\documentclass[12pt,preprint]{aastex}
\usepackage{ulem} 

\slugcomment{Accepted by ApJ on March 13, 2009.}

\shorttitle{M4 WD Cooling Sequence}
\shortauthors{Bedin et al.}

\begin{document}

\def\subr #1{_{{\rm #1}}}


\title{The End of the White Dwarf Cooling Sequence in M4:\ \\~  an efficient
approach\footnote{ Based on observations with the NASA/ESA {\it Hubble 
Space Telescope}, obtained at the Space Telescope Science Institute,
which is operated by AURA, Inc., under NASA contract NAS 5-26555,
under GO-10146.}}

\author{ Luigi R.\  Bedin\altaffilmark{2},  
         Maurizio Salaris\altaffilmark{3}, 
         Giampaolo Piotto\altaffilmark{4},  
         Jay     Anderson\altaffilmark{2}, 
         Ivan R.\    King\altaffilmark{5},
         Santi    Cassisi\altaffilmark{6}. 
         }

\altaffiltext{2}{Space Telescope Science Institute, 3800 San Martin 
Drive, Baltimore, MD 21218; [bedin;jayander]@stsci.edu}

\altaffiltext{3}{Astrophysics Research Institute, Liverpool John Moores
University, 12 Quays House, Birkenhead, CH41 1LD, UK; ms@astro.livjm.ac.uk}

\altaffiltext{4}{Dipartimento di Astronomia, Universit\`a di Padova,
Vicolo dell'Osservatorio 2, I-35122 Padova, Italy;
giampaolo.piotto@unipd.it}

\altaffiltext{5}{Department of Astronomy, University of Washington,
Box 351580, Seattle, WA 98195-1580; king@astro.washington.edu}

\altaffiltext{6}{INAF-Osservatorio Astronomico di Collurania,
via M. Maggini, 64100 Teramo, Italy;
cassisi@oa-teramo.inaf.it}

\begin{abstract}
We  use  14  orbits of  ACS  observations  to  reach  the end  of  the
white-dwarf  cooling  sequence  in   the  globular  cluster  M4.   Our
photometry  and completeness  tests show  that the  end is  located at
magnitude  $m_{\rm   F606W}=28.5\pm0.1$,  which  implies   an  age  of
$11.6\pm0.6$ Gyr (internal errors  only).  This is consistent with the
age from fits to the main sequence turn-off ($12.0\pm1.4$ Gyr).
\end{abstract}


%%%%%%%%%%%%%%%%%%%%%%
%
\section{Introduction}
%
%%%%%%%%%%%%%%%%%%%%%%

Stellar evolution theory predicts that the remnants of the majority of
stars with masses less  than $\sim$6--7$M_{\odot}$ become white dwarfs
(WDs) after their  progenitors have lost a large  part of the envelope
during the asymptotic giant branch (AGB) phase.
Single-star  evolution WDs  are  compact objects  with masses  between
$\sim$0.5 and $\sim$1.0 $M_{\odot}$ (increasing slowly with increasing
progenitor mass),  comprising an electron-degenerate  core made almost
exclusively of carbon and  oxygen, surrounded by a thin non-degenerate
envelope ($\sim$1\% in mass fraction).  The temperature of the CO core
is  too  low to  ignite  nuclear  reactions,  and electron  degeneracy
prevents any  sizable contraction and  associated temperature increase
that would  ignite carbon  burning.  The energy  radiated from  the WD
photosphere is therefore the thermal energy of the non-degenerate ions
contained in  the core,  and the  WD evolution can  be described  as a
cooling process (Mestel 1952).

WD  evolutionary  tracks in  the  color--magnitude  diagram (CMD)  are
constant-radius sequences,  which the stars  follow as they  fade with
time to progressively fainter magnitudes and redder colors.
Given that the older the WD  the fainter its magnitude, the CMD of the
WD population  in a star  cluster is expected  to show a cut-off  at a
certain magnitude, because  no WD has had time  to become fainter than
some limiting brightness (see, e.g., Winget et al.\ 1987).

From  a theoretical point  of view,  one can  predict a  WD luminosity
function (LF) for  a single-burst stellar population of  given age and
metallicity  by  assuming  an  initial-mass-final-mass  relation,  and
adopting  an  initial  mass  function (IMF),  along  with  appropriate
lifetimes for the progenitors (see, e.g., Richer et al.\ 2000).

All theoretical WD LFs for old  stellar populations show a peak at the
faint  end, caused  by  the ``piling  up''  at the  bottom  of the  WD
sequence,  as the cooling  rate slows  down with  decreasing effective
temperature (and luminosity).
The  peak  moves  towards  fainter  luminosities as  the  cluster  age
increases,  reflecting the  increasing age.   For simplicity,  we will
hereafter refer to this peak in the WD LF as the end of the WD CS. The
location of the peak is therefore a measure of the age, and it is only
mildly  affected  by the  adopted  IMF  (see,  e.g., Prada  Moroni  \&
Straniero 2007, Salaris 2009).
In  summary, it  is  possible to  use  the bottom  of  the WD  cooling
sequence (WD CS)  to obtain an estimate of  a cluster age, independent
of the main-sequence turn-off (MS TO).

Theoretical predictions of the cut-off brightness as a function of age
rely on accurate WD cooling models.
Despite  the  apparent  simplicity  of  the  structure  of  WD  stars,
theoretical WD cooling times  still show sizable uncertainties.  These
are related  first of all to  the chemical stratification  of the core
and mass fractions of the non-degenerate He and H layers.
Additional uncertainties stem from the equation of state of the dense,
cool ion gas,  and the treatment of the  crystallization process.  The
opacity of the  envelope and the surface boundary  conditions are also
critical for the determination of  the cooling rate of faint WDs older
than a few Gyr, and  are still subject to non-negligible uncertainties
(Montgomery et al.\ 1999, Salaris et al.\ 2000, Salaris 2009).

Thus  far, age estimates  that made  use of  the entire  WD CS  in old
clusters  (older than  $\sim$6 Gyr)  have been  performed in  only two
cases:\
(1) the  metal-poor globular cluster  NGC 6397  (Richer et  al.\ 2006,
Anderson et al.\ 2008a),  and (2) the old super-solar-metallicity open
cluster NGC 6791 (Bedin et al.\ 2008a).  In the case of NGC 6397 there
seems to  be a remarkable agreement  between the age  derived from WDs
and the  age derived from MS TO  stars (Hansen et al.\  2007).  In the
case  of NGC  6791, by  contrast, there  seems to  be a  conflict with
results from standard WD models (Bedin et al.\ 2005a, 2008a, 2008b).

M4 (NGC 6121) is the closest  globular cluster to the Sun, but because
of the large extinction along its line of sight, its apparent distance
modulus (12.5 $\pm$ 0.1 mag in $V$, Richer et al.\ 2004) is comparable
to that of NGC 6397 (12.6 $\pm$  0.1 mag in $V$, Richer et al.\ 2008).
The metallicity  of M4 is intermediate  between those of  NGC 6397 and
NGC 6791, making it particularly  useful to test the ages derived from
the WD CS.

The  observation  of  the piling-up  in  the  WD  LF  in M4  has  been
previously attempted with 123 orbits of {\it HST} using the Wide Field
and   Planetary  Camera   2  (WFPC2),   by  Richer   et   al.\  (2002,
GO-8679). Their  photometry did indeed  reach the sharp rise  near the
bottom of the WD LF, but failed to detect a clear drop beyond the peak
(Richer  et al.\  2004, Hansen  et al.\  2002, 2004).   These authors,
however, were  nevertheless able  to obtain an  age estimate  from the
cooling   sequence  by   fitting  theoretical   models  to   the  full
two-dimensional WD color-magnitude diagram.

Thanks to the superior capabilities of the Advanced Camera for Surveys
(ACS),  and thanks also  to our  improved astrometric  and photometric
techniques, we can now push the limits somewhat deeper, and detect the
drop of the WD LF, in just 14 {\it HST\/} orbits.

In  Section  2  we  give  a brief  description  of  the  observations,
reduction  procedures,  and  calibrations.  Section  3  describes  the
selection process of real stars and well-measured stars, and Section 4
deals with the completeness. In  Section 5 we describe how we obtained
proper motions, to produce a decontaminated (but incomplete) sample of
the best-measured stars.   In Section 6 we describe  the observed CMD,
and  in  Section  7 we  present  the  WD  LF.   Section 8  presents  a
comparison between the observed CMD of M4 with that of NGC 6397, while
Section  9 gives a  detailed comparison  of observations  with theory.
Finally, we give a summary of the results in Section 10.

%%%%%%%%%%%%%%%%%%%%%%%%%%%%%%%%%%%%%%%
%
\section{Observations and Measurements}
%
%%%%%%%%%%%%%%%%%%%%%%%%%%%%%%%%%%%%%%%

%%%
\subsection{Observational Data}
%%%

All data were taken with the  Wide Field Channel (WFC) of the ACS. The
observations in the F606W filter consist of 20 exposures of $\sim$1200
s (10 HST  orbits in total) collected between July  and August 2004 as
part  of the  program  GO-10146 (PI:  Bedin).   The observations  were
divided  evenly between  two  different orientations  so  as to  avoid
putting PSF artifacts  in the same relative positions  with respect to
real  stars.   The  F775W  filter  data  consist  of  4  exposures  of
$\sim$1200 s (2 HST orbits) taken on  2005 June 27 as part of the same
GO program, and 10 exposures of  360 s (2 HST orbits) obtained in 2002
July   and  2003   June  as   part   of  the   program  GO-9578   (PI:
Rhodes). GO-9578  had additional data  from another HST orbit  in 2003
January, but due  to extremely high background this  data was unusable
for our purposes.
Hereafter we will  refer to all of these F606W  and F775W exposures as
the {\it deep} images.  All images collected within GO-10146 were well
dithered (according to the precepts  of Anderson \& King 2000), and in
{\sf LOW SKY} mode.

We also took  a sample of short exposures to link  the bright stars on
the horizontal branch and near  the main sequence turn-off to the same
photometric system. At the beginning  of each of our first five visits
we took one short exposure, as follows:\  0.5 s $+$ 10 s in F606W, and
0.5  s $+$  2$\times$  10 s  in  F775W.  In  addition,  we used  F606W
archival material:\  1 s +  2 $\times$  25 s +  2 $\times$ 30  s (from
program GO-10775, PI:\ Sarajedini).

%%%%%%%%%%%%%%%%%%%%%%%%%%%%%%%%%%%%%%%
%
\subsection{Measurements and reduction}
\label{PHO}
%
%%%%%%%%%%%%%%%%%%%%%%%%%%%%%%%%%%%%%%%

The photometry  was carried out  with the software tools  described in
great detail by Anderson et al.\ (2008b).
Briefly, the method consists of a two-pass procedure.  We start with a
pair  of  ``Library  PSFs'',  one  for each  filter,  which  had  been
constructed  from other  data sets.   Each library  PSF consists  of a
9$\times$10 array  of PSFs that  represent how the ACS/WFC  PSF varies
across the detectors which we perturb slightly to fit each exposure as
described  in  Anderson  \&  King  (2006, AK06).   We  then  use  this
perturbed PSF  to measure the  fluxes and positions of  the relatively
isolated  and relatively  bright stars,  using the  code  described in
AK06.  We  then adopt the first  deep F606W exposure  as the reference
frame,  and  find  the  6-parameter  linear  transformation  from  the
distortion-corrected frame of each exposure into this frame, using the
positions of common  stars.  We also put the  magnitudes into a common
zero-point system based on the deepest F606W and F775W exposures.

In  the   second  pass  a  sophisticated   software  routine  analyzed
simultaneously  all the  individual  exposures to  find the  brightest
sources and to measure a single position and a F606W and an F775W flux
for each of them.
It then subtracts  out the newly found sources  from each exposure and
iterates the finding procedures  until no more significant objects are
found.  To qualify  as found, an object has to be  detected as a local
maximum  in at  least 6  out of  the 34  deep images.   Many  of these
detections are simply coincident noise peaks, but we include them here
to demonstrate how distinct  the background-noise distribution is from
our  sources.  Anderson  et al.\  (2008b) describes  how we  manage to
avoid including PSF artifacts in our star lists.

In  addition   to  solving  for   positions  and  fluxes   during  the
simultaneous-fitting process  above, we also compute  a very important
image-shape diagnostic  parameter for each source:\ RADXS.   This is a
measure of  how much flux there is  in the pixels just  outside of the
core, in excess of the prediction  from the PSF.  (We measure it using
the pixels between $r=1.0$ and $r=2.5$, and it is reported relative to
the star's  total flux.)  RADXS is  positive if the  object is broader
than the  PSF, and  negative if  it is sharper.   This quantity  is of
great help  in distinguishing between stars and  galaxies whose images
are nearly as sharp; the latter are especially numerous in the part of
the CMD where the faint white dwarfs lie.

\begin{figure*}[ht!!]
\epsscale{1.00}
\plotone{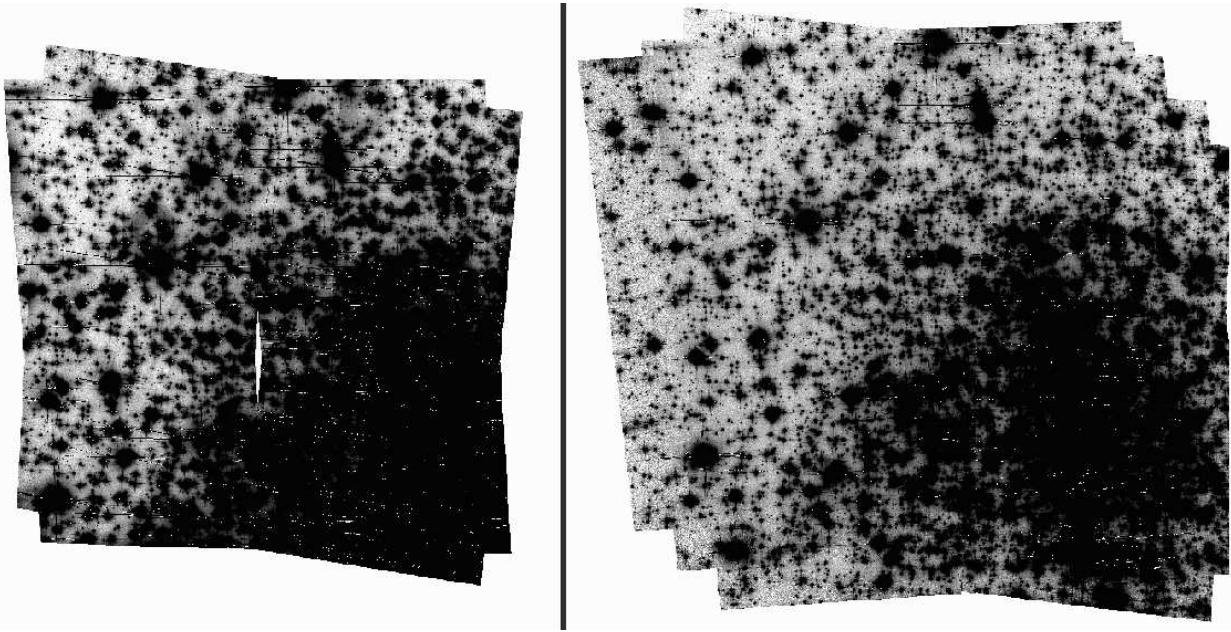}
\caption{
$(Left):$ Stacked image of deep exposures in filter F606W. 
$(Right):$ Stacked image of deep exposures in filter F775W. 
It is immediately clear that the region where we will be able to reach
the  faintest magnitudes  will  necessarily exclude  the  core of  the
cluster, where the halos of  bright stars hide faint objects. The core
regions  instead provide  good  statistics for  the fast  evolutionary
phases of  the bright WDs. [Note  that the brightest WDs  are close to
saturation in F606W].
}
\label{ssi}
\end{figure*}

We  compute a  stacked image  free of  cosmic-rays and  artifacts (see
Fig.~\ref{ssi}).  We  use this stack to construct  two other important
parameters  for  each  detected  object:\  these  are  the  local  sky
background  (SKY) and its  r.m.s.\ deviation  (rmsSKY), as  defined in
Bedin et al.\ (2008a).

Artificial-star  (AS) tests  are  performed using  the same  procedure
(Anderson  et  al.\ 2008b).   For  each AS,  a  position  and a  F606W
magnitude  are chosen in  a random  way; the  F814W magnitude  is then
chosen so that  the star's color puts  it on the ridge line  of the WD
sequence (drawn  by eye).  The AS  is added into each  exposure at the
appropriate location, in the form  of an appropriately scaled PSF with
Poisson noise. The software routine then operates blindly, finding and
measuring all the stars.  We  examine the resulting list of sources to
determine whether  the artificial star was  recovered.  The artificial
stars are  used not only as  a measure of the  completeness; they also
serve, at a  number of stages of the procedure,  to help in developing
and calibrating our criteria for the choice of valid stars.

For  the  analysis, we  will  follow  closely  the procedure  recently
applied with success to the study of the the WD CS of the open cluster
NGC 6791 (Bedin et al.\ 2008a).

%%%
\subsection{Input-output corrections}
%%%

\begin{figure}[ht!!]
\epsscale{1.00}
\plotone{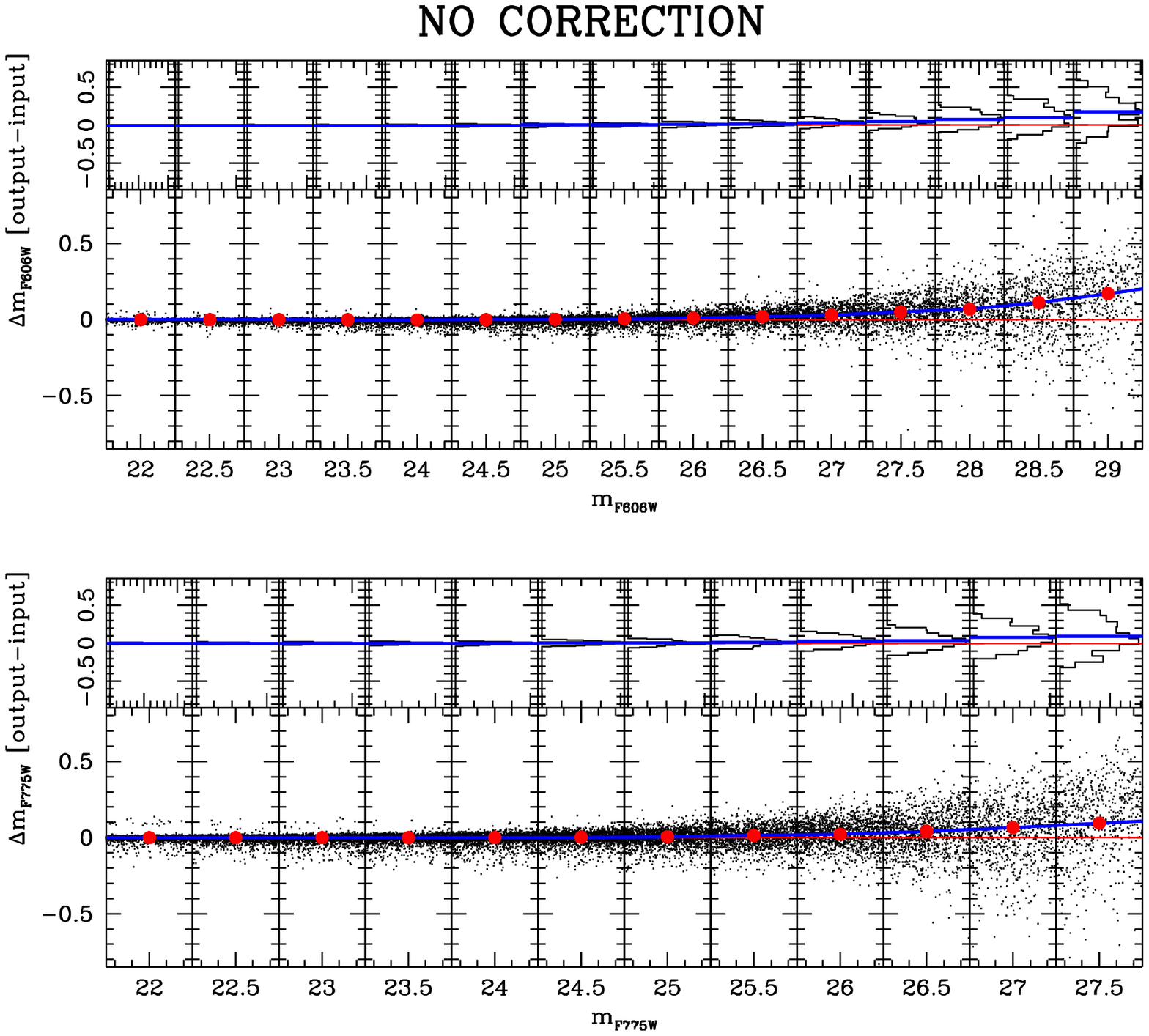}
\caption{
Difference  between inserted  and  recovered AS  magnitudes, for  each
filter separately (top panel F606W, bottom panel F775W).
The lower part  of each panel shows data points,  while the upper part
gives histograms of the same data.
In  the upper  parts the  red lines  indicate the  position of  a null
difference,  while the  blue  lines  are the  observed  median in  the
corresponding  magnitude  interval.   In   the  lower  parts  the  red
horizontal  lines indicate again  the position  of a  null difference,
while the  blue lines  now are the  splines through the  median values
obtained in each magnitude interval (red dots).}
\label{CORR1}
\end{figure}

%
Artificial-star tests allow  us to assess precisely how  many stars at
each magnitude  level were  recovered by our  algorithms and  how well
they  were  recovered.   Figure~\ref{CORR1}  displays  the  difference
between the inserted and  recovered magnitudes of artificial stars for
both filters.  We divided the magnitude range covered by ASs into bins
of  half a magnitude,  and estimated  the median  of the  (output {\it
minus} input) values for each of  the stars within the bin (red dots).
Magnitudes of the very faintest  stars in the field are systematically
overestimated by as much as $\sim$0.2 magnitude at $m_{\rm F606W}=29$.

\begin{figure}[ht!!]
\epsscale{1.00}
\plotone{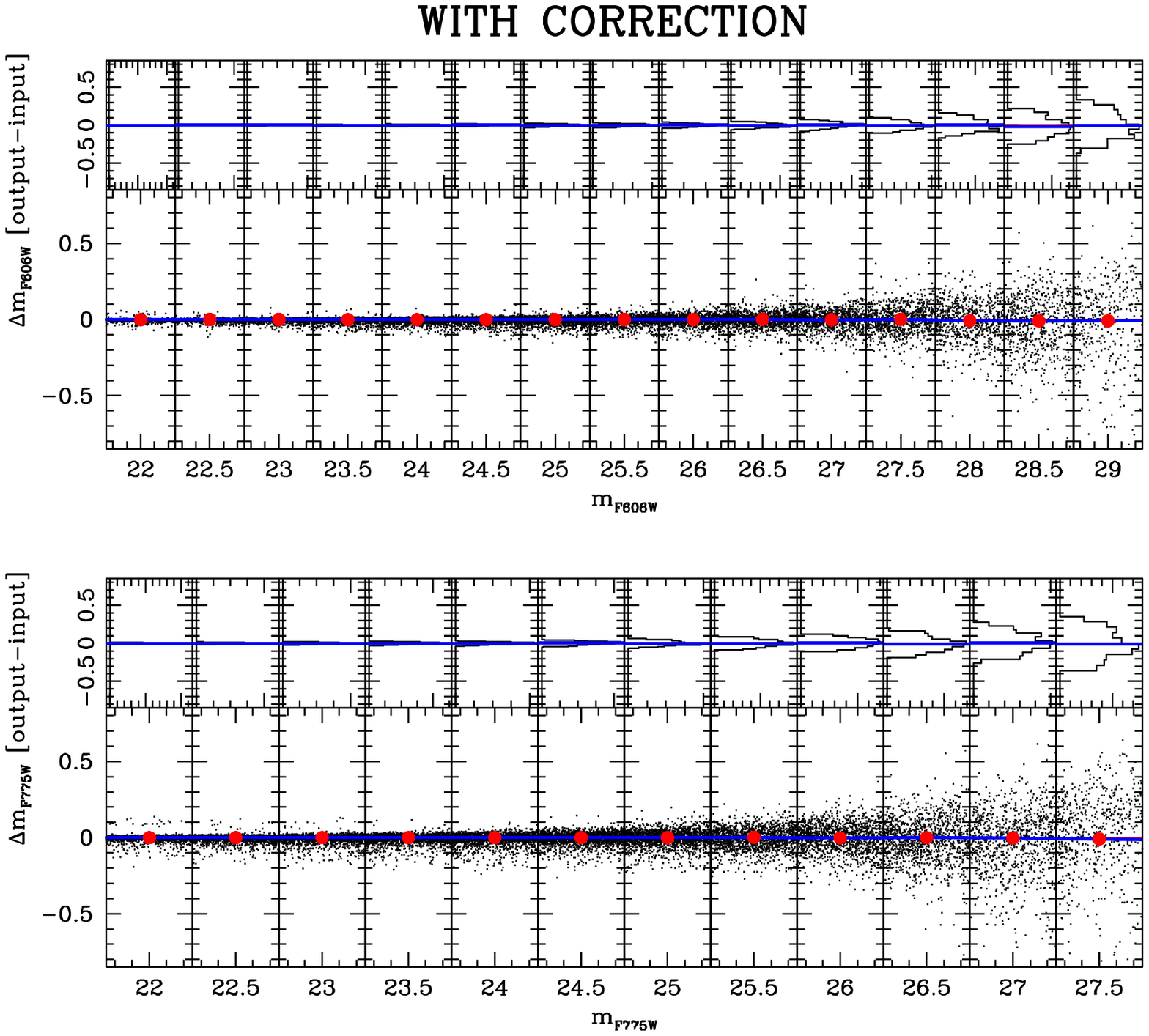}
\caption{  
Difference  between inserted  and  recovered AS  magnitudes after  the
photometric-offset correction was applied.}
\label{CORR2}
\end{figure}

These were used  as fiducial points for our  empirical correction.  We
applied  to  each  star   (artificial  or  real),  in  each  magnitude
independently, the  correction obtained by a  quadratic spline through
the derived  fiducial points.  Figure~\ref{CORR2} shows  the result of
the correction, which seems to  do a good job overall.  Hereafter, all
the  magnitudes  will be  corrected  for  this systematic  photometric
offset.

%%%
\subsection{Photometric zero points}
%%%

We calibrate  the photometry to the WFC/ACS  Vega-mag system following
the procedure given in Bedin  et al.\ (2005b), and using the encircled
energy and the zero points given by Sirianni et al.\ (2005). Hereafter
we will refer to these  calibrated magnitudes with the symbols $m_{\rm
F606W}$, and $m_{\rm F775W}$.

%%%
\section{Selection of Stars}
%%%

\begin{figure*}[ht!!]
\epsscale{1.00}
\plotone{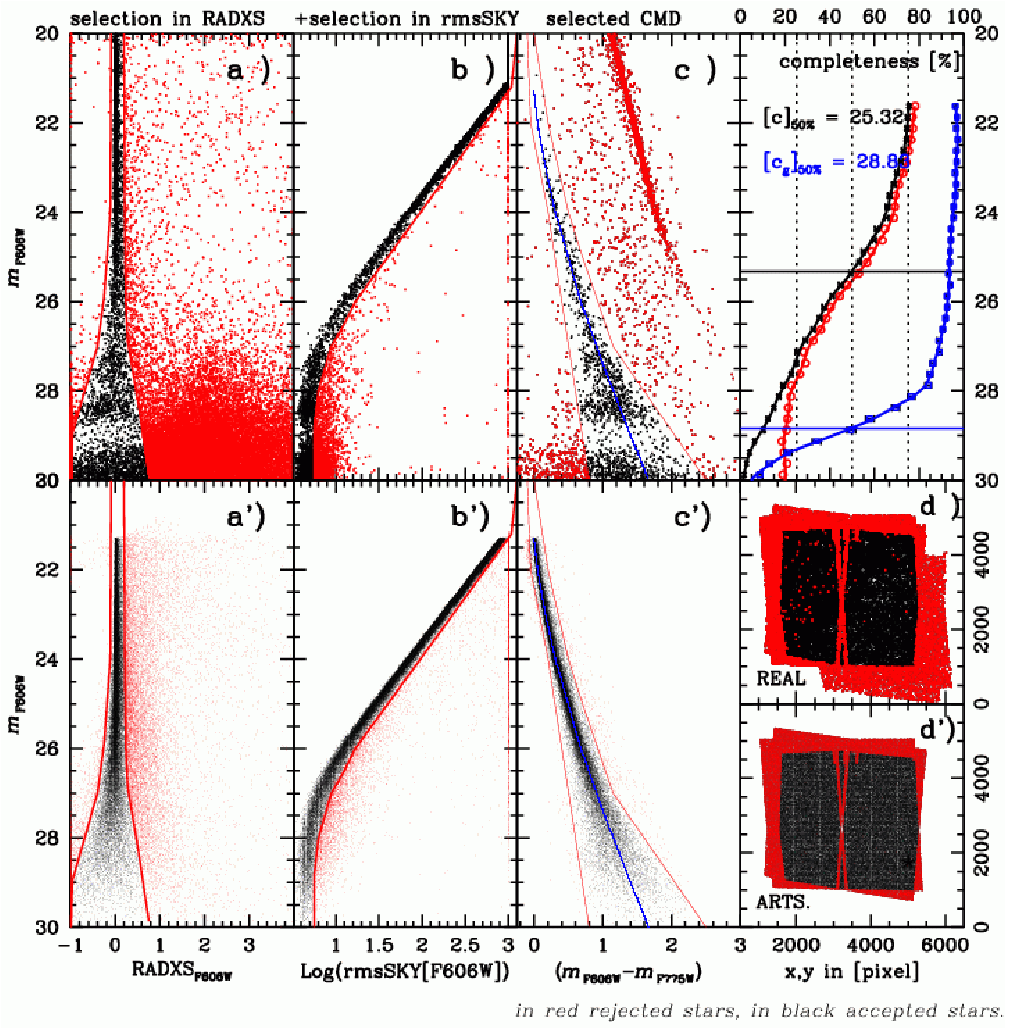}
\caption{
{\it  Panels~a), a$^\prime$):} The  parameter RADXS  as a  function of
magnitude in  filter F606W, for  real and AS.  Detections  between the
two red lines are considered to be real stars.
{\it Panels~b),  b$^\prime$):} The parameter  rmsSKY as a  function of
magnitude after selection in RADXS, for real and AS.
{\it Panels~c),  c$^\prime$):} the  CMD for the  selected real  and AS
detections.  Note  that everything below  $m_{\rm F606W}\simeq29.5$ is
essentially noise (see text). All the objects between the two thin red
lines are considered WDs.
{\it Panels~d), d$^\prime$):} spatial  distribution of the real and AS
detections.  In black,  objects that fell in all  20 F606W deep images
and in at least 10 of the F775W deep images.
{\it Top right}: The  black crosses show the conventional completeness
$c$,  while   the  blue   squares  are  the   low-rmsSKY  completeness
$c\subr{g}$ that is defined in the text.  The red circles indicate the
fraction of the image area $f_g$ where the bumpiness of the of the sky
offers no impediment to finding a star, at each magnitude.
Note that at the faintest  magnitudes about $\sim$20\% of the observed
field is useful for determination of the WD LF.
}
\label{all}
\end{figure*}

It is particularly important to identify galaxies and remove them from
the star lists, because many of them fall in the part of the CMD where
the faint  end of the WD  sequence lies.  To  distinguish these barely
resolved objects from point sources, we use the shape parameter RADXS,
which measures any excess flux just outside of the PSF core.  Here the
artificial stars  are very instructive,  since they show us  how RADXS
should behave with magnitude for  true point sources.  Panels $a)$ and
$a')$  in Fig.~\ref{all}  show us  the  trend of  RADXS against  F606W
magnitude for artificial stars (bottom),  and for objects that we have
thus far  considered to be real (top).   It is clear that  many of the
objects that we have thus far included as stars are more extended than
the artificial stars, which  by definition have truly stellar profiles
(because  we made them  that way).   We use  the distributions  of the
artificial stars to mark out  boundaries in these diagrams that should
retain nearly all the objects that are truly stars; they are indicated
by the red lines, which were drawn  in such a way as to include almost
all  of the  recovered stars.  They  are repeated  identically in  the
corresponding real-star panel.   Note that the tail of  objects on the
right side in  the AS panels is produced by  star-star blends that our
simultaneous-fitting  routine  was  not  able  to  separate  into  two
components.  These should certainly  be eliminated from our photometry
lists.

%
%echo $( 34.044  -2.5*LG(  {1,2,3,...5} * (sqrt(290)/sqrt(20-1))/.2) )
%
It is also  clear that most of the objects  found fainter than $m_{\rm
F606W}\simeq29.5$ are just peaks of the noise in the sky fluctuations.
For reference, the  lowest sky value in the  typical 1200s F606W image
is $\sim$290  DN, with a  corresponding Poisson noise of  $\sim$17 DN,
which averaged on 20  deep F606W images becomes $\sim$17/$\sqrt(20-1)$
$\simeq$  4 DN.   Our  PSF models  tell  us that  the brightest  pixel
contains $\sim$20\% of  the light, so the total  collected flux of the
(PSF-fitted) peak of noise would be $\sim$20 DN, i.e.\ an instrumental
magnitude of $-2.5 \times \log_{10}(20) = -3.2$.  Since our zero point
in F606W,  $ZP_{\rm F606W}$, is  34.04, our $1\sigma$ level  in filter
F606W is $m_{\rm F606W}\simeq30.8$,  the $3\sigma$ level is at $\simeq
29.6$, and the $5\sigma$ level at $\simeq 29.0$.
Because  of  the many  uncertainties  in  these  calculations, in  the
following we  will consider as  {\it not significant} all  the sources
that our code measures to  be fainter than $m_{\rm F606W}=29$.  It is,
however, instructive  to show them  in our plots during  the selection
procedures.

\begin{figure*}[ht!!]
\epsscale{1.00}
\plotone{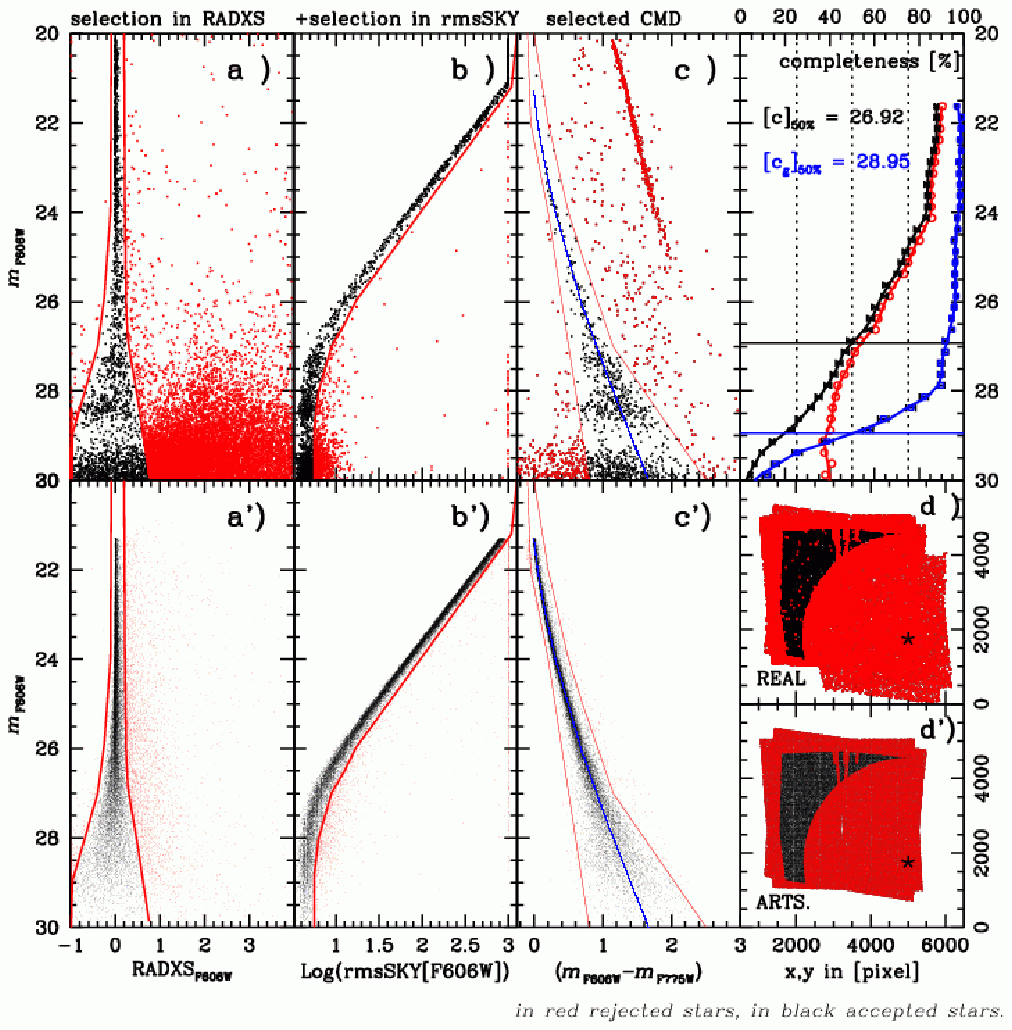}
\caption{
As in Figure~\ref{all}, but selecting  only those objects which are at
least $140''$ from the assumed position for the cluster center (marked
with a $\star$) in panels {\it d)} and {\it d$^\prime$)}.
}
\label{best}
\end{figure*}

Our second image-quality parameter, rmsSKY,  has a minor impact in the
selection  of stars,  but it  gives us  invaluable information  on how
suitable the surroundings of each star are for good photometry.
As the profiles  of brighter stars have higher  counts in their wings,
and are  consequently affected by greater Poisson  noise, their rmsSKY
also increases.
This effect is clearly seen in panels $b)$ and $b')$ of Fig~\ref{all},
which displays rmsSKY  for real and artificial stars  in a way exactly
analogous to panels $a)$ and $a')$ for RADXS.  Again, limit lines have
been drawn  by eye for the  artificial stars, in a  way that separates
the stars found on flat background from those found in noisier areas.
We will  use rmsSKY  again in Section  \ref{compl}, where it  plays an
even more important role.

Panels  $c)$ and $c')$  of Fig.~\ref{all}  show the  CMD for  real and
artificial stars  after applying our selection criteria  on both RADXS
and rmsSKY.  The blue line  shows the chosen fiducial line along which
the ASs were  added.  We will hereafter consider as WDs  of M4 all the
stars that fall between the two red lines.

As an additional condition we selected only those ASs and real sources
that fell  in portions  of the sky  covered by  all 20 deep  images in
F606W  and  at  least 10  of  the  deep  F775W images.  These  spatial
selections are  shown in panels  $d)$ and $d')$ of  Fig.~\ref{all} (in
red we show the objects that do not pass these spatial criteria).

%%%%%%%%%%%%

The WD  CS that we find  here for M4 is  qualitatively consistent with
the  one exhibited  by  Hansen et  al.\  (2004) based  on their  WFPC2
observations in a different field  (see Fig.~4 in Richer et al. 2004),
although it  is difficult to compare  the photometry in  the F775W and
F814W filters quantitatively.

%%%%%%%%%%%%%%%%%%%%%%%
%
\section{Completeness}
\label{compl}
%
%%%%%%%%%%%%%%%%%%%%%%%

Bedin et al.\ (2008a) make the  point in their Section 4 that the main
impediment  to finding  faint stars  in {\it  HST} images  of globular
clusters stems not from  near-neighbor crowding, but from the presence
of a few extremely bright giants. These giants have extended PSF halos
that create a  mottled background, making it hard  to find faint stars
in their vicinity.   But in between these bright stars,  it is easy to
find and measure faint stars well.   Whereas a given faint star may be
found in only 25\%  of the field as a whole, if  we search only in the
flat areas between the bright giants,  we may find it over 75\% of the
time.

We found  in Bedin et  al.\ (2008a) that  our ability to find  a given
faint star  depends most  strongly on the  flatness of  the background
upon which it  is inserted.  If we can determine  what fraction of the
image has  a favorable  background, we can  then reliably  assess what
fraction of such stars could be found.  This is the motivation for our
construction of the rmsSKY parameter: it allows us to search for stars
only where they can be found easily.  This has two advantages: when we
find them, we  measure them more accurately and,  in addition, we tend
to have  far fewer  false detections, since  such detections  are more
likely in the less flat portions of the image.

Our completeness corrections for faint stars thus have two sources: an
incompletness due  to the region  not searched, and  the incompletness
within the region searched.  The  net completness is $c$, the fraction
of stars recovered and it is the product of $f_g$ (the fraction of the
image that  is searchable) and  $c_g$ (the fraction of  the searchable
region  where  the  source  was  found).  We  define  $f_g$  for  each
brightness level of star to be the fraction of the the image where the
background is flat  enough to allow the artificial  stars to be found.
Both factors  come from  artificial-star tests, as  shown in  Bedin et
al.\ (2008a).

The top  right panel of  Fig.~\ref{all} shows the  completeness levels
for our field.
Figure~\ref{best}  is  the  same  as  Fig.~\ref{all},  but  explicitly
considers only the very best part (lowest crowding) of the whole field
that we studied,  the part of the image that is  more than 2800 pixels
($\simeq140^{\prime\prime}$) from the estimated cluster center [marked
with a $\star$ in panels $d)$ and $d')$].
We will explore the end of the WD CS only in the flatter parts of this
sub-region of our  field, but we will use the  whole field to increase
the  star-count statistics  in  the fast  evolutionary  phases of  the
bright part of the WD CS.

In Fig.~\ref{best}  the 50\%  level of {\it  ``overall completeness''}
$c$  is at  $m_{\rm F606W}=26.92$,  while  that for  the {\it  ``local
completeness''} $c\subr{g}$  it is  at $m_{\rm F606W}=28.95$.   Had we
followed conventional  rule of trusting only corrections  with $c$ $>$
50\%,  we would  have lost  the faintest  $\sim$2 magnitudes  and thus
missed  the  cutoff   of  the  WDCS,  which  the   real  CMD  and  the
artificial-star CMD clearly show we have reached.  The top right panel
of Figs.~\ref{all} and \ref{best} also shows the fraction of the image
area  suitable for  detection  of faint  sources  (open circles).  The
product  of  this curve  and  $c\subr{g}$  is  of course  the  overall
completeness $c$.  Note how the fraction of usable field becomes about
$\sim$20\%  at the  faintest  magnitudes in  Fig.~\ref{all} (or  about
$\sim$40\%  in  Fig.~\ref{best}),  which   is  crucial  to  the  large
correction applied to the observed WD LF (see Fig.~\ref{wdlf}).

%%%%%%%%%%%%%%%%%%%%%%%
%
\section{Proper motions}
\label{PMS}
%
%%%%%%%%%%%%%%%%%%%%%%%

Proper  motions  were  measured  with  a  technique  similar  to  that
described in detail in Bedin et al.\ (2003, 2006).
As our  first-epoch data (hereafter,  epoch I) we used  5$\times$360 s
F775W images collected  on July 14, 2002 (GO-9578),  while as epoch II
we used  4$\times$1200 s  exposures taken in  June 27, 2005  (from our
GO-10146),  also   in  filter  F775W  to  avoid   the  possibility  of
color-dependent  systematic errors.   To compute  a proper  motion for
each  star, we  first measured  a mean  position for  the star  in the
reference frame  at each  epoch, using all  the exposures  within that
epoch.  The proper motion, then, is the difference between the second-
and first-epoch  positions, divided by  the time baseline  (2.95 yrs),
and multiplied  by the pixel scale  (49.7 mas/pixel, van  der Marel et
al.\  2007).  The  reference  frame is  aligned  to have  $x$ and  $y$
parallel to RA  and Dec, respectively, so the proper  motions are in a
properly aligned frame  as well.  The zero point of  the motion is the
cluster's bulk motion, since the reference frame was defined by member
stars.

Thanks to the very large proper  motion of M4 with respect to both the
field and  background objects [about $\sim$15 mas  yr$^{-1}$ (Bedin et
al.\ 2003), i.e., almost a whole  pixel in three years], as long as we
were able to  measure even a crude position in both  epochs we were to
get  a good  discriminant between  field objects  and  cluster members
(down to $m_{\rm F606W}\sim$29).

Our first epoch is considerably shallower than the second epoch, which
means that there is a good  chance that the faintest stars will not be
found in the first epoch and we  will not get motions for them.  If we
were to  require that all stars  in our final sample  have good proper
motions, then our completeness would  be severely limited by the first
epoch.  Nevertheless, even though  we cannot do quantitative work with
the proper  motions, we  can still  use them in  a qualitative  way to
illustrate the properties of the stars that we could measure.

\begin{figure*}[ht!]
\epsscale{1.00}
\plotone{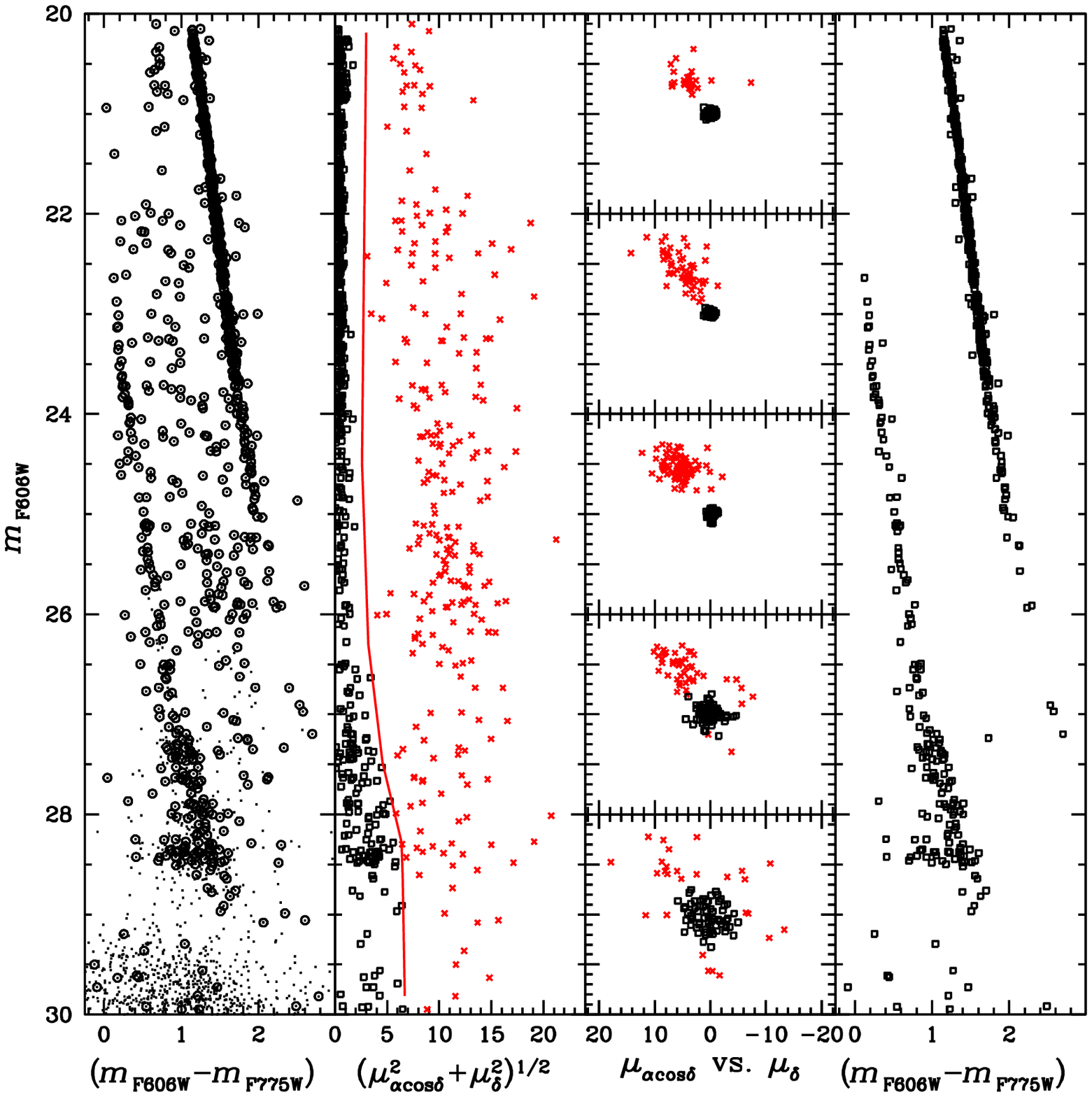}
\caption{
{\it (Left):} Color-magnitude diagram for  all stars in panel {\it c)}
of Fig.~\ref{best}.  Small dots  are all the stars, circles highlights
the stars  for which  it was  possible to measure  a position  in both
epochs.
{\it (Second panel from left): } Total proper motion, in mas yr$^{-1}$
relative to  the mean proper  motion of the cluster.   Errors increase
towards fainter  magnitudes.  The line  was drawn in order  to isolate
the objects that have  member-like motions (squares); red crosses show
objects that we consider non-members.
{\it  (Third  panel from  left):  }  Vector  point diagrams  for  five
different magnitude intervals indicated on the left-hand scale.
$(Right):$  Proper  motion-selected  CMD  (using  criteria  in  second
panel).
}
\label{pms}
\end{figure*}

The  left panel of  Fig.~\ref{pms} shows  as small  dots all  stars in
panel  $c)$ of  Fig.~\ref{best}. Stars  for which  it was  possible to
estimate a  proper motion are  highlighted with open circles.   In the
second panel we show for this subsample of stars the magnitudes of the
proper motions.  Cluster  stars have an internal dispersion  of few km
s$^{-1}$ (Peterson,  Rees \& Cudworth~1995), which at  the distance of
M4  corresponds  to less  than  $\sim$0.5  mas  yr$^{-1}$.  All  field
objects (Galactic disk in the foreground, and bulge stars, halo stars,
and galaxies  in the background)  have a very different  proper motion
distribution, which clearly stands out  from that of the members. Note
the increasing size of random errors with decreasing S/N.  In order to
separate members from non-members we drew the red line by eye.
Non-members,  to  the right  of  the red  line,  are  marked with  red
crosses.   In the third  panel, using  the same  symbols, we  show the
vector-point diagrams (in mas  yr$^{-1}$) for five different magnitude
bins ($m_{\rm F606W}=$20--22; 22--24; 24--26; 26--28, and 28--30).
The CMD that  results from the cleaning criterion  of the second panel
is shown in the rightmost panel of the same figure.

\begin{figure*}[ht!]
\epsscale{1.00}
\plotone{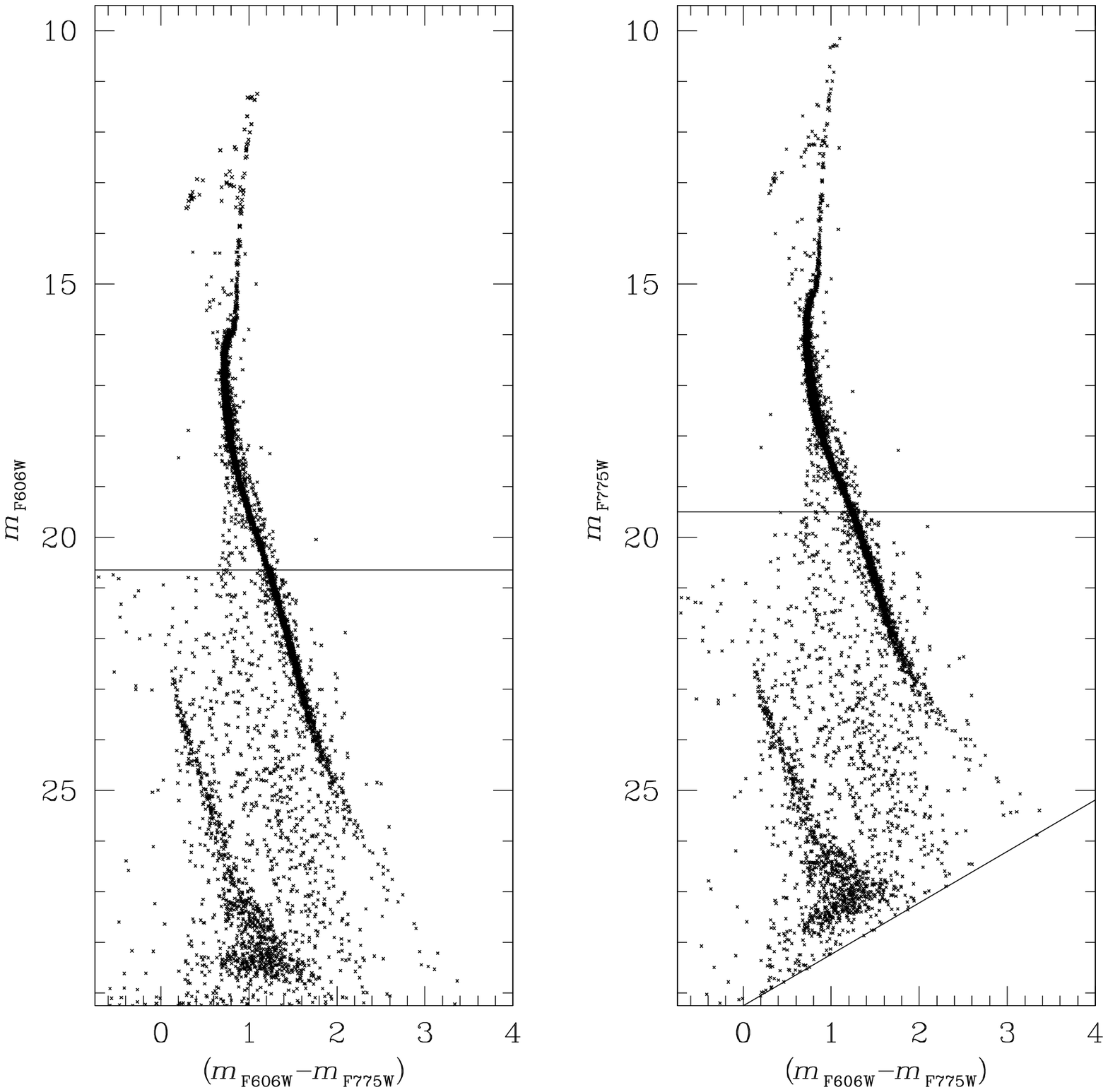}
\caption{
The objects in these CMDs  (not selected for proper motions) span more
than a factor 10$^7$ in luminosity.  The bright stars were measured in
the  short exposures,  then adjusted  to the  zero point  of  the deep
exposures.  The horizontal lines  mark the saturation levels of images
with exposure times of $\sim$1200  s.  For deep photometry we show the
stars in panel $c)$ of Fig.~\ref{all}.
In the right-hand  panel we show only the same  objects that appear in
left-panel.  A diagonal line marks  the location of the cut introduced
by the faint $m_{\rm F606W}$ limit.
}
\label{cmd}
\end{figure*}

%%%%%%%%%%%%%%%%%%%%%%%%%%%%%%%%%%%%%%%%%%%%%%
%
\section{The Color-Magnitude Diagram}
%
%%%%%%%%%%%%%%%%%%%%%%%%%%%%%%%%%%%%%%%%%%%%%%

In Fig.~\ref{cmd} we present  the entire color-magnitude diagrams, for
deep  and  short  exposures,  in both  observational  planes  ($m_{\rm
F606W}-m_{\rm   F775W},m_{\rm  F606W}$)  and   ($m_{\rm  F606W}-m_{\rm
F775W},m_{\rm F775W}$).  These CMDs span  a factor of more than $10^7$
in luminosity.
All the  main evolutionary sequences  are very well defined,  from the
region of the  hydrogen-burning limit up the MS  through the turn-off,
then almost to the tip of the red giant branch.  The horizontal branch
and white dwarf cooling sequence are also well represented.

This CMD, complete in all its  relevant sequences, will be used in the
following in order to measure  the distance, reddening, and age of the
cluster, from  both the MS turn-off and  the WD CS.  A  tricky part in
producing the CMD of Fig.\ 7 was to link the $\sim$1200 s exposures to
the  relatively short  exposures  ($\sim$10 s)  in  filter F606W.   To
perform this task we used the  calibrated catalog of M4 by Anderson et
al.\  (2008b)  to  extend  to  brighter objects  our  calibrated  deep
photometry.  Unsaturated stars that  are common to the two independent
calibrations  show that  they are  in  agreement to  better than  0.01
magnitude.
Our short  exposures were zero-pointed consistently to  the stars that
are in common with the Anderson et al.\ (2008b) catalog.
In filter F775W  there were no difficulties in  putting the photometry
of  the short  exposures  on the  zero  point of  the deep  calibrated
exposures, as a  number of stars are well exposed  in both the several
360 s and the 1200 s frames.

The photometry obtained from short exposures comes from the first pass
described  in  Section  2,   wherein  starlists  are  made  from  each
individual exposure and these lists merged to yield the list of common
stars.   This   yields  different  selection   effects  and  therefore
different  completenesses than  the comprehensive  2nd-pass procedure.
But since these stars are all  quite bright, it is safe to assume that
they  are  nearly  100\%  complete  and suffer  very  little  neighbor
contamination.  We mark  with a horizontal line the  level below which
photometry comes from deep exposures.

Even  if the  CMD is  subject to  incompleteness, the  observed  WD CS
clearly piles up at  $m_{\rm F606W}=28.5\pm0.1$, followed by an abrupt
drop in  a magnitude range where  our completeness is  still high.  In
the following Section we will give a more quantitative analysis of the
completeness of  the star counts in the  lower part of the  WD CS, and
properly estimate the position of the WD LF peak.

\begin{figure*}[ht!]
\epsscale{1.00}
\plotone{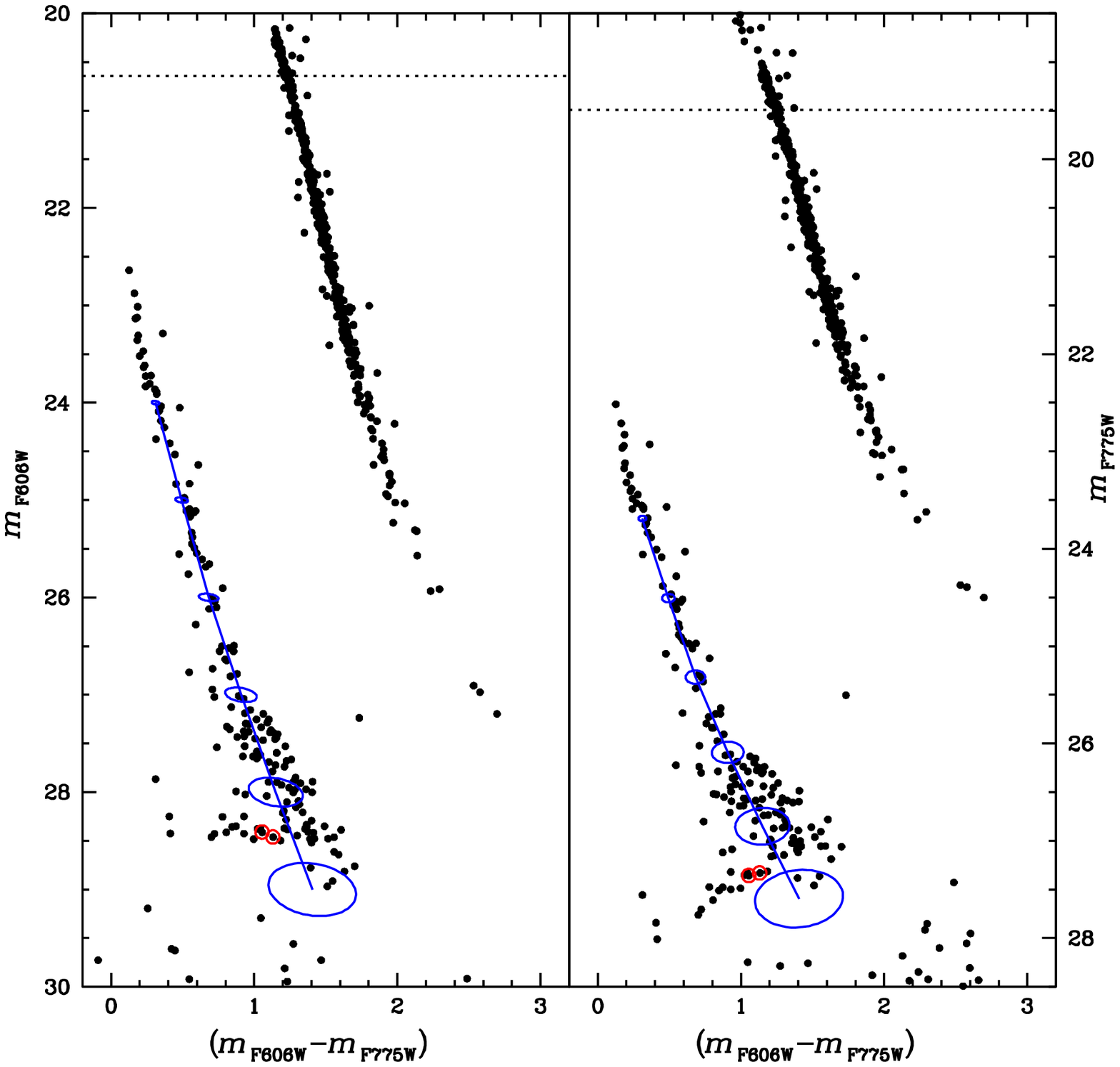}
\caption{
Decontaminated  CMDs  for a  sub-sample  of  stars  for which  it  was
possible  to  estimate proper  motions.  Dotted  lines indicate  where
saturations set in.  Error ellipses along the WD fiducial line on each
panel indicate the estimate of photometric rms errors along the WD CS.
Two stars  on the WD LF peak  are highlighted by red  open circles and
shown in the stack images, in Fig.~\ref{bts}.
}
\label{corr}
\end{figure*}

In   Figure~\ref{corr}   we  show   the   two   CMDs   for  just   the
proper-motion-decontaminated subsample  obtained in Section~\ref{PMS},
along with the  photometric error ellipses along the  WD fiducial line
for six magnitude levels (at $m_{\rm  F606W}=24$, 25 , 26, 27, 28, and
29).

\begin{figure*}[ht!]
\epsscale{1.00}
\plotone{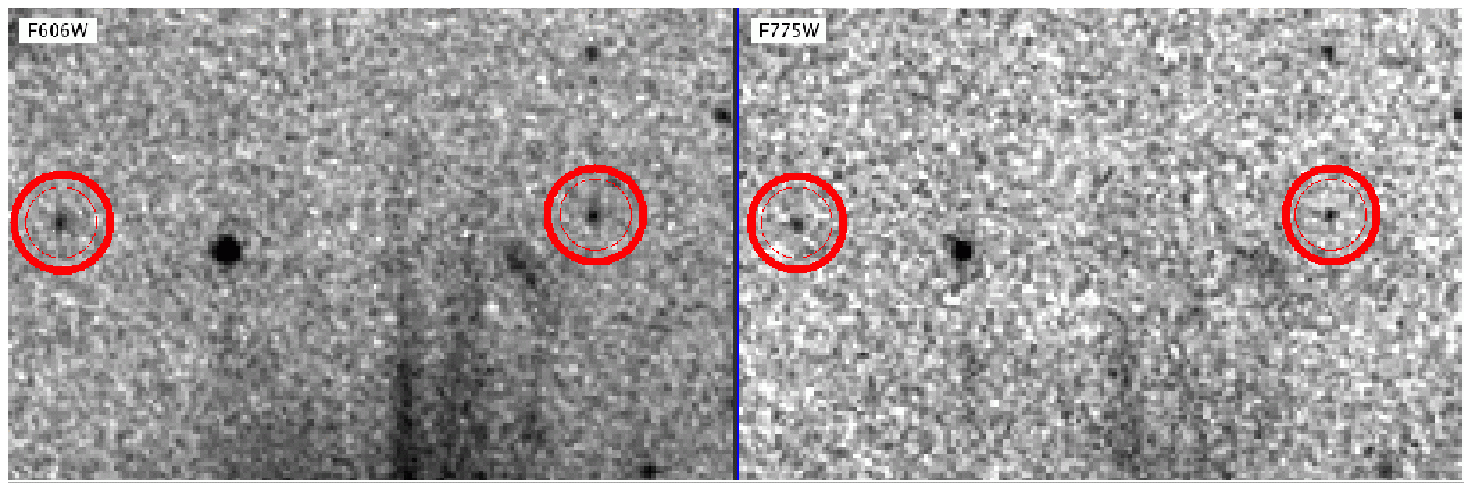}
\caption{
The two  stars highlighted with red  circles on the  stacked images in
filter F606W (left) and F775W (right)  are stars that lie at the faint
end of the WD CS.
These two  objects are real  and relatively easy to  measure [$(m_{\rm
F606W}-m_{\rm F775W},m_{\rm  F606W}) =  (1.13,28.46)$ for the  star on
the left, and (1.06,28.41) for the one on the right].
}
\label{bts}
\end{figure*}

Although  these CMDs  are even  more incomplete,  and not  amenable to
reliable  correction,  they  include  the  best  astro-photometrically
measured stars in our sample.
To  give  a  better idea  of  the  objects  we  are dealing  with,  in
Fig.~\ref{bts} we  show that two  stars in the  peak of the WD  LF are
clearly visible and measurable in both stacked images.

At the bottom of the WD CS there is some hint of a blue turn. However,
the  photometric errors  are too  large to  enable us  to  confirm its
presence and measure its extension.

%
% !!!!! IMPORTANT !!!!! DOUBLE CHECK AT PROOF STAGE THAT THIS FIGURE COMES OUT RIGHT. 
%

%%%%%%%%%%%%%%%%%%%%%%%%%%%%%%%%%%%%%%%%%%%%%%
%
\section{The White Dwarf Luminosity Function}
%
%%%%%%%%%%%%%%%%%%%%%%%%%%%%%%%%%%%%%%%%%%%%%%
\label{WDLF}

Without a proper-motion elimination of field stars, the best we can do
to derive a white dwarf luminosity function (WD LF) is what we did for
NGC 6791 in Bedin et al.\ (2008a):\ we will use the RADXS parameter to
remove as many galaxies and artifacts  as we can, and delimit in color
the region of  the CMD in which we consider stars  to be white dwarfs.
For both  of these tasks  the artificial-star experiments prove  to be
extremely useful.

First,  panel  $a')$ of  Fig.~\ref{all}  shows  our  treatment of  the
parameter RADXS.   We used  the results  of our AS  tests to  draw our
discriminating  lines,  in  such  a  way  as  to  include  nearly  all
identified objects;  we then applied those same  selection criteria to
the real stars, in panel $a)$.

We then chose our boundaries in  the CMD, as indicated by the thin red
lines in panel $c')$ of Fig.~\ref{all}.  The boundaries were chosen so
that only a marginal fraction of the artificial stars were lost, while
the boundaries are set broadly  enough to include all stars that might
be  white dwarfs  of some  sort---including WD  binaries,  and unusual
types of WDs.

\begin{figure*}
\epsscale{1.00}
\plotone{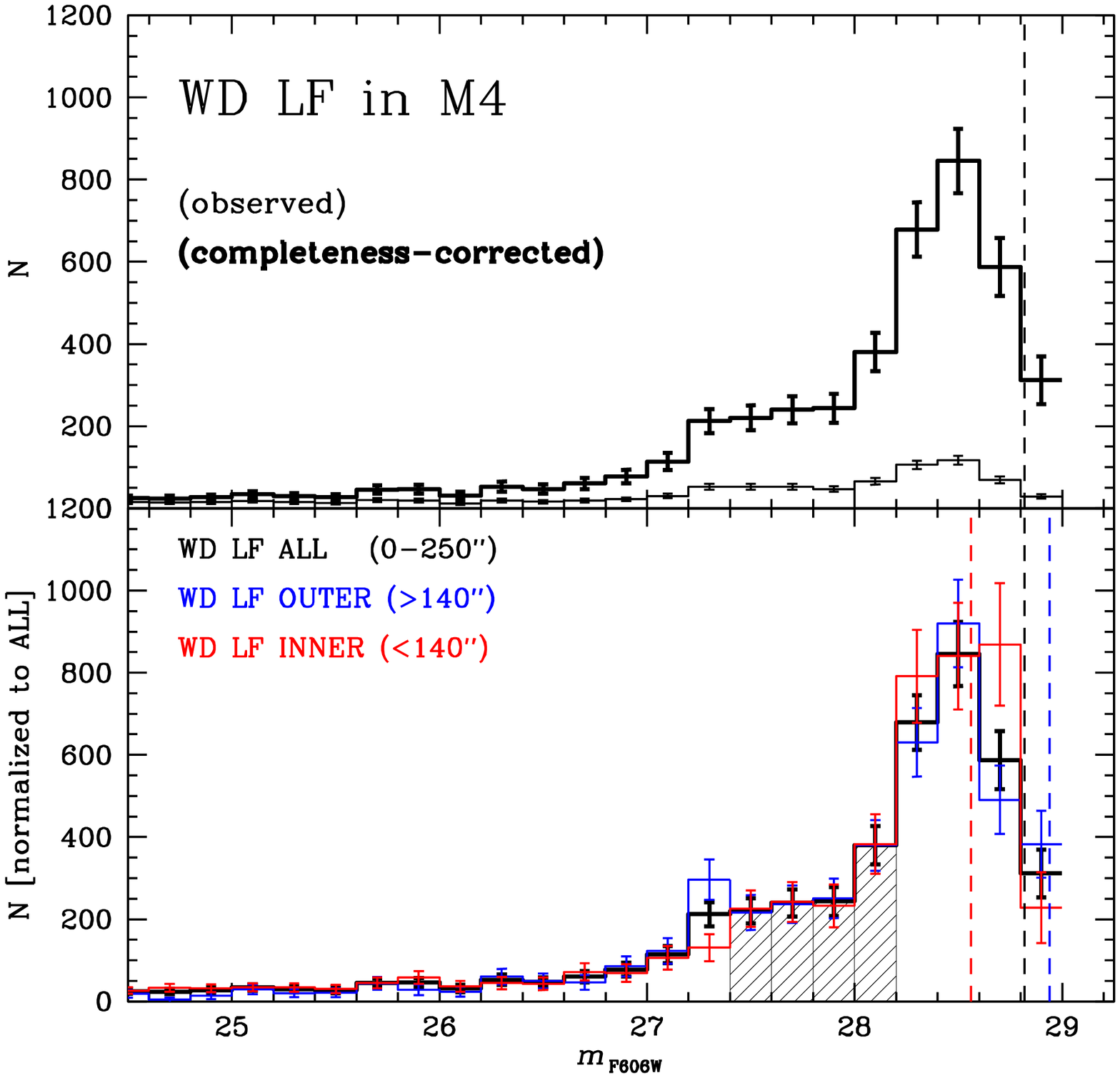}
\caption{ 
$(Top:)$  White dwarf luminosity  function in  our field,  as observed
(thin line),  and after correction for completeness  (thick line). The
vertical dashed  line indicates the  level of $c_g=50\%$,  below which
our completeness becomes unreliable.
$(Bottom:)$ Completeness-corrected  LF for  WDs within $140''$  of the
cluster  center (INNER,  in  red),  and for  WDs  at larger  distances
(OUTER,  in blue), with  the $c_g=50\%$-completeness  magnitude levels
marked by vertical dashed lines of the corresponding color.
The shaded bins of the histograms indicate where the normalization was
done.
Note  that  the  $c_g=50\%$  level  for  OUTER  reaches  the  faintest
magnitudes, allowing us to follow the drop of the WD LF for almost two
bins after the peak at $m_{\rm F606W}=28.5$.
Note also that crowding seems to  make the WD LF peak somewhat broader
in the INNER sub-sample than in the OUTER.
}
\label{wdlf}
\end{figure*}

In the  top panel of Fig.~\ref{wdlf} we  show the WD LF  in our entire
field  as observed  (thin-line  histogram), and  after correction  for
completeness   (thick-line  histogram).   The  vertical   dashed  line
indicates  the level  of  $c_g$=50$\%$, below  which our  completeness
becomes unreliable.
The bottom  panel of the same figure  shows the completeness-corrected
LF for WDs within $140''$ from the cluster center (labeled with INNER,
in red), and for WDs outside that limit (OUTER, in blue).
The shaded bins of the  histograms indicate where the normalization to
the entire field WD LF (ALL) was done.
Note   that  $c_g$=50$\%$-level   for  OUTER   reaches   the  faintest
magnitudes, allowing us to follow the drop of the WD LF for almost two
bins after the peak at $m_{\rm F606W}=28.5$.
%

%%%-%%%

The WD  LF in the entire field  (see top panel of  Fig.~10) displays a
clear  increase  in  the   number  of  objects  fainter  than  $m_{\rm
F606W}\sim$27.3, which corresponds to the location where $T_{\rm eff}$
decreases below  $\sim$ 5000~K, assuming  the distance modulus  and WD
isochrones described  in Section~9.   This feature is  consistent with
the  luminosity  function jump  found  by  Hansen  et al.\  (2004)  at
$V>27.4$ in their data, which they attribute to a drop in photospheric
hydrogen  opacity and  a  change  in the  boundary  conditions of  the
cooling models.

\begin{table*}[ht!]
\caption{
Completeness-corrected  white dwarf  luminosity function  (ALL).  Note
that the last  point, at $m_{\rm F606W}=$28.9, is  beyond the limit of
reliability for $c_g$.
}
\begin{tabular}[h]{ccc|ccc|ccc} 
\hline\hline
& & & & & & & & \\
$m_{\rm F606W}$ & $N_c$ & $\sigma_{N_c}$ & 
$m_{\rm F606W}$ & $N_c$ & $\sigma_{N_c}$ & 
$m_{\rm F606W}$ & $N_c\,$ & $\sigma_{N_c}$\\  
& & & & & & & & \\
\hline
%#Nx     Nyc  sNyc  Nx     Nyc   sNyc   Nx       Nyc   sNyc
%#
 20.1 & 0.00 & 0.00 & 23.1 &  4.27 &  2.46 & 26.1 &  31.50 &  9.09 \\
 20.3 & 0.00 & 0.00 & 23.3 & 14.50 &  4.59 & 26.3 &  52.90 & 12.14 \\
 20.5 & 0.00 & 0.00 & 23.5 & 14.80 &  4.68 & 26.5 &  47.09 & 11.77 \\
 20.7 & 0.00 & 0.00 & 23.7 & 22.48 &  5.80 & 26.7 &  60.36 & 13.85 \\
 20.9 & 1.32 & 1.32 & 23.9 & 22.69 &  5.86 & 26.9 &  77.36 & 16.49 \\
 21.1 & 0.00 & 0.00 & 24.1 & 13.93 &  4.64 & 27.1 & 113.98 & 20.81 \\
 21.3 & 0.00 & 0.00 & 24.3 & 17.46 &  5.27 & 27.3 & 212.41 & 29.18 \\
 21.5 & 1.33 & 1.33 & 24.5 & 24.82 &  6.41 & 27.5 & 220.12 & 30.53 \\
 21.7 & 1.33 & 1.33 & 24.7 & 24.22 &  6.47 & 27.7 & 239.84 & 33.26 \\
 21.9 & 1.33 & 1.33 & 24.9 & 26.92 &  6.95 & 27.9 & 243.40 & 35.50 \\
 22.1 & 1.34 & 1.34 & 25.1 & 33.99 &  8.01 & 28.1 & 380.57 & 46.85 \\
 22.3 & 1.35 & 1.35 & 25.3 & 29.79 &  7.69 & 28.3 & 678.79 & 65.93 \\
 22.5 & 1.36 & 1.36 & 25.5 & 27.50 &  7.63 & 28.5 & 845.50 & 78.17 \\
 22.7 & 5.50 & 2.75 & 25.7 & 45.70 & 10.22 & 28.7 & 587.60 & 70.74 \\
 22.9 & 8.40 & 3.43 & 25.9 & 46.78 & 10.73 & 
{\it 28.9} & {\it 311.59} & {\it 57.86} \\
\hline 
\hline
\label{WDLFtab}
\end{tabular}
\end{table*}

The WDLF ends at a magnitude where our completeness is still reliable.
In Table~\ref{WDLFtab} we  report the values of the  entire WD LF.  We
must  treat  these numbers  with  care,  however.   Because of  energy
equipartition the heavier  WDs might have sunk into  the center of the
cluster, where we are considerably more incomplete.  This might result
in biases on the relative numbers of the WD LF.

Also,  our numbers  include  no correction  for point-like  background
galaxies  that may  have been  included in  our WD  counts,  though we
expect this contamination  to be small because of the  care we took in
eliminating non-stellar objects  (see Section 3).  We now  turn to the
question of contamination by background objects.

%%%%%%%%%%%%%%%%%%%%%%%%%%%%%%%%%%%%%%%%%%%%%%
%
\subsection{Contamination by background galaxies}
\label{hudf}
%
%%%%%%%%%%%%%%%%%%%%%%%%%%%%%%%%%%%%%%%%%%%%%%

Following  the recipe  and  the  motivations given  by  Bedin et  al.\
(2008a) for  a similar study of NGC  6791, in this section  we use the
Hubble  Ultra Deep  Field  (HUDF;  Beckwith et  al.\  2006) to  assess
whether the contamination  by point-like background galaxies seriously
affects our WD CS, and/or WD LF.

We are thankful for two  happy circumstances.  First, part of the HUDF
data-base is in the  same photometric system ($m_{\rm F606W}$, $m_{\rm
F775W}$)  as our  program, and  second,  we have  already reduced  and
calibrated  the   HUDF  with  the  same  algorithms   (Bedin  et  al.\
2008a). The reduction of the  HUDF subset of images presented in Bedin
et al.\  (2008a) is  complete down to  the magnitudes of  interest for
this paper (once appropriate reddening is added).
We note also  that the HUDF, also observed with  ACS/WFC, has the same
area as our field.

In  the left-hand  panel of  Fig.~\ref{hudf} we  show the  CMD  of the
objects  in  the  HUDF   that  survived  our  selection  criteria  (in
particular the  selection on RADXS).  A triangle  indicates the region
occupied by the bulk of point-like objects in the HUDF.
The bottom  edge of  the triangle was  chosen at an  arbitrarily faint
magnitude  below  which objects  are  no  longer  relevant to  our  M4
observations (see below).

If we assume that the distribution of the field objects in the HUDF is
the same  as that of  the objects  in our M4  field, we can  apply the
corresponding shifts in  reddening ($A_{\rm F606W}-A_{\rm F775W}$) and
absorption ($A_{\rm F606W}$)  and see if these objects  overlap any of
the location of the WS CS.
In the right-hand panel of  Fig.~\ref{hudf} we apply these shifts both
to the HUDF objects and to  the triangle, using the reddening law that
applies to  cool objects (4,000 K),  from Bedin et  al.\ (2005b, Table
3).
We also  indicate with  a dashed line  the position that  the triangle
would have if  we had used the extinction law  for hot objects (40,000
K), from Table 4 of the same work.
As we can  see, very few of the transformed  HUDF objects fall between
the two thin red lines that we used to select WDs in Section 3.
In fact, this  is a generous upper limit  to the contamination, since,
as  we  have explained  in  Sect. \ref{compl},  we  are  using only  a
fraction of our M4 field to go deep.  (Fig.~\ref{best}).
The  black  dots  in  right  panel of  Fig.~\ref{hudf}  show  the  CMD
presented in Fig.~\ref{best}, while the blue circles are the subsample
of proper motion selected stars shown in Fig.~\ref{pms}.

\begin{figure*}[ht!]
\epsscale{1.00}
\plotone{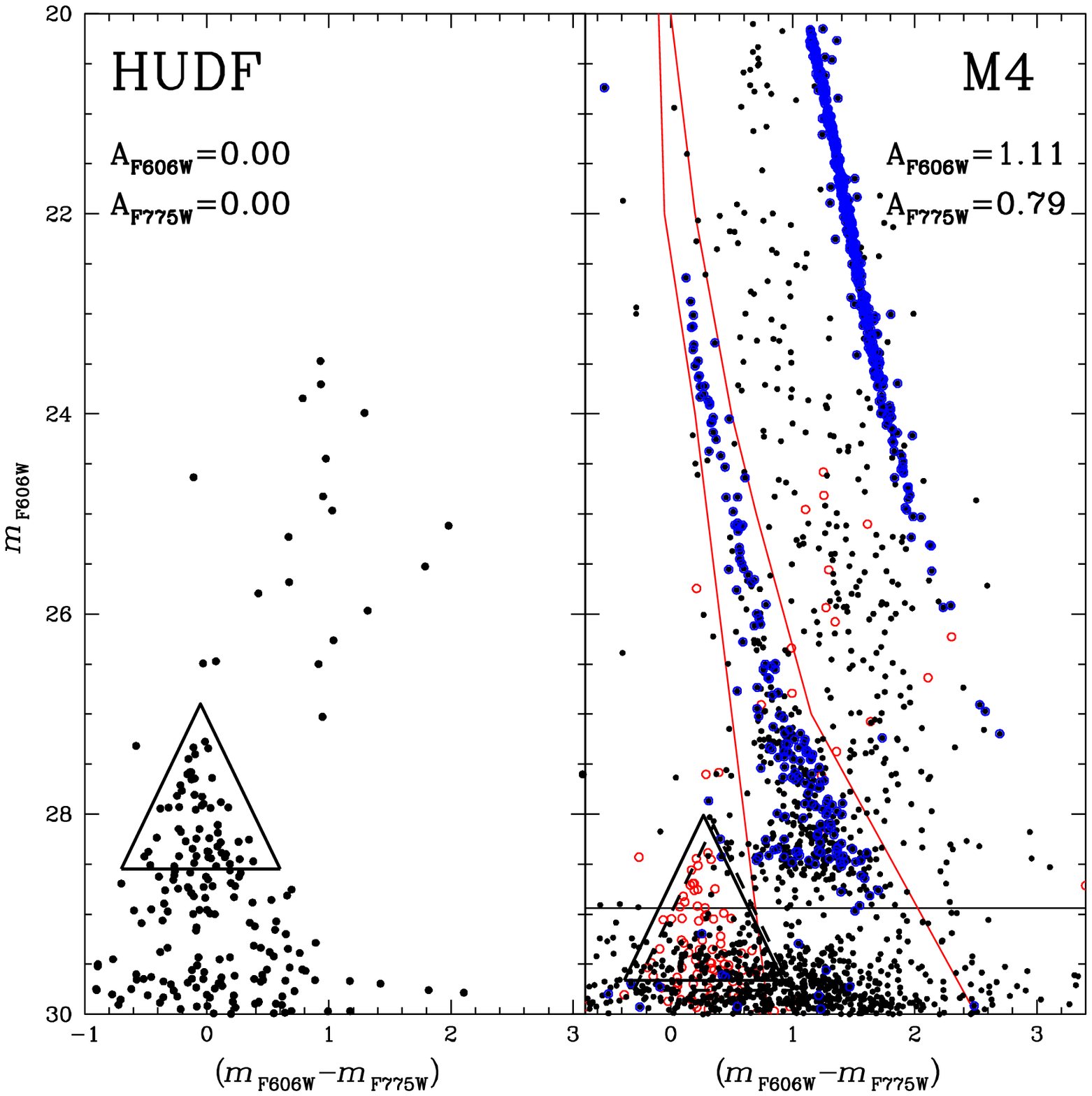}
\caption{
$(Left):$ CMD  obtained from a subset  of the Hubble  Ultra Deep Field
images, in  its native observational plane. A  triangle highlights the
location of  the point-like background sources in  the magnitude range
that might affect our WD counts.
$(Right):$ Open red circles show HUDF objects shifted according to the
estimated  reddening in  our M4  field. The  triangle is  also shifted
according to the  value of the extinction for  cool objects, estimated
using  Table 3  in Bedin  et al.\  (2005b).  The  dashed  triangle has
instead had extinction  added for hot objects according  to Table 4 in
Bedin et al.\  (2005b).  The two thin red lines  delimit the region of
what  we  consider   a  WD.   The  horizontal  line   marks  the  50\%
completeness level of  $c_g$. Black points mark objects  in panel $c)$
of Fig.~\ref{best}, while the blue circles highlight the proper-motion
selected sample of Fig.~6.
}
\label{hudf}
\end{figure*}

%
%%%%%%%%%%%%%%%%%%%%%%%%%%%%%%%%%%%%%%%%%%%%%%%%%%
\section{Comparison with NGC 6397 {\textit HST} Observations}
%%%%%%%%%%%%%%%%%%%%%%%%%%%%%%%%%%%%%%%%%%%%%%%%%%

In the previous sections, we showed that with only 14 {\sl HST} orbits
we are able to  reach the bottom of the WD CS.  This result might seem
surprising  in view  of the  much larger  number of  orbits previously
dedicated to the  detection of the end  of the WD CS in  M4 (Richer et
al.\ 2002, 2004,  Hansen et al.\ 2002, 2004), and  in NGC 6397 (Richer
et al.\ 2006, 2008, Hansen et al.\ 2007, and Anderson et al.\ 2008a).
As for M4,  the papers based on the GO-8679 123  {\sl HST} orbits used
only  WFPC2  images, and  the  superior  quality  of ACS/WFC  is  well
known. As for  the spectacular results of Richer  and collaborators on
NGC 6397, they are indeed based on the 126 orbits of GO-10424.
In addition, the F775W filter has a somewhat lower efficiency than the
F814W filter  used in  GO-10424, and  our field is  in a  more crowded
region of the cluster.
We believe  that a major contributor  is our overall  strategy of data
reduction and analysis.
In fact, in this section we will show that the end of the WD CS in NGC
6397 can  indeed be reached using  just a subsample of  14 orbits from
GO-10424. The purpose of this exercise is to validate the results that
we will present in Section~\ref{the}.

The  subsample of  GO-10424 data  was extracted  merely by  taking the
first images of that dataset,  in alphabetic order, until we summed up
a total observing time equivalent to  our total exposure time in M4 in
F606W ($\sim$24,000 s), and in F775W ($\sim$8,500 s).
To reach these  totals, we selected the first 34  images in F606W, and
the first 12  F814W images of GO-10424, each  with an average exposure
time of  $\sim$710 s ($\pm$70 s).   (Note that to have  the same total
observing time  used for  M4, we  need to use  15 orbits  of GO-10424,
instead of 14.)

\begin{figure*}[ht!]
\epsscale{1.0}
%\epsscale{2.30}
\plotone{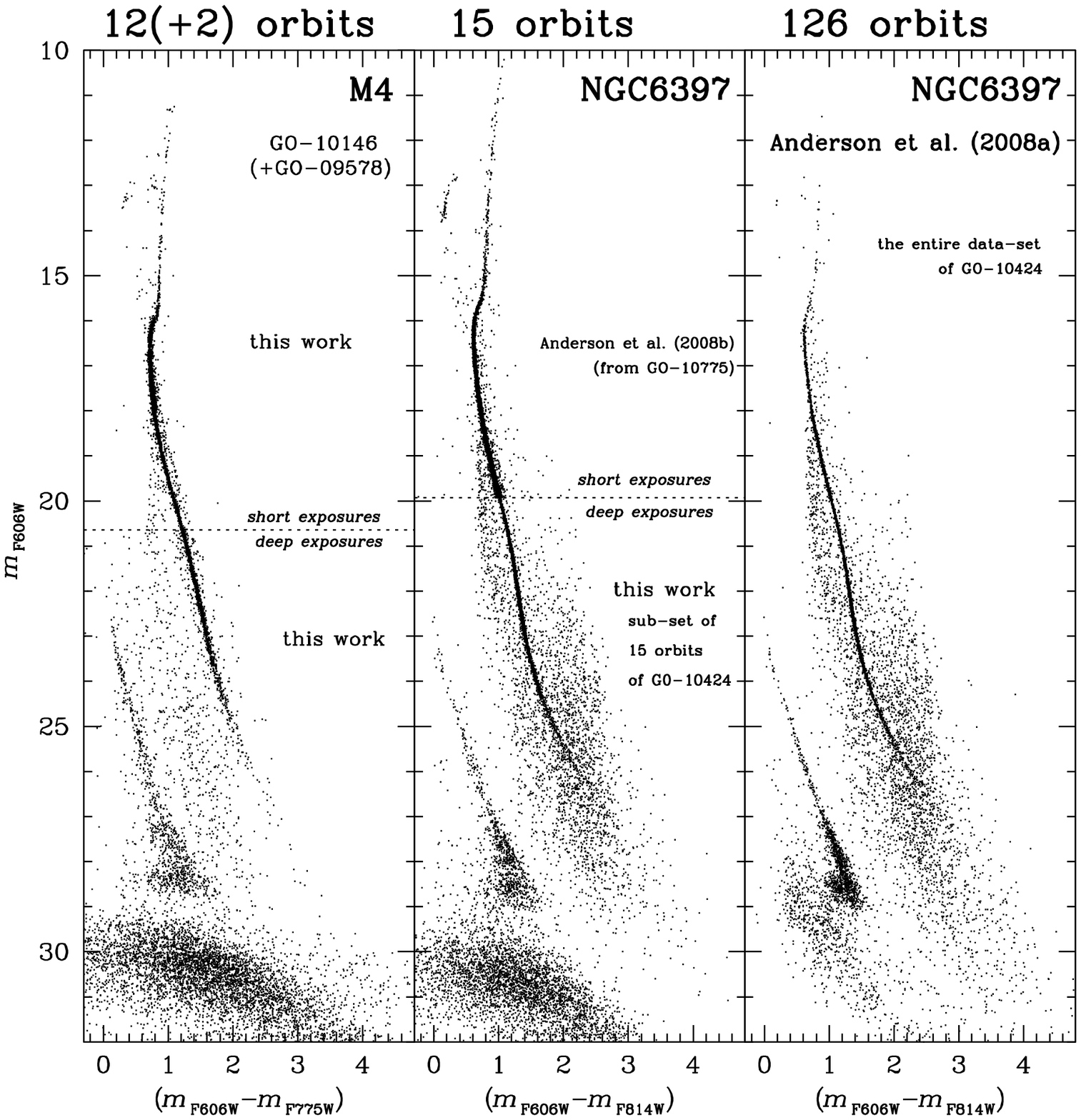}
\caption{
$(Left):$ CMD  for M4 obtained  combining our 12-orbit program  with 2
orbits from the archive.
Dotted lines indicate the onset of saturation for deep exposures.
$(Middle):$ CMD  for NGC  6397 using a  subset of the  GO-10424 images
having the same  total amount of exposure time for  each filter of our
M4 program.   Photometry above saturation  comes from a  central field
analyzed by Anderson et al.\ (2008b).
$(Right):$ CMD for NGC 6397 obtained by Anderson et al.\ (2008a) using
the entire 126-orbit data set.
}
\label{compa}
\end{figure*}

In Figure~\ref{compa} we compare our CMD  of M4 (left) with the one of
NGC 6397 that was obtained with  an equal exposure time, and using the
same procedures described in Sections 2 and 3 (middle panel).
In the right-hand panel we show  the CMD of NGC 6397 obtained from the
entire GO-10424 data set (126 orbits), as obtained by Anderson et al.\
(2008a).

Note that: (1) we can reliably detect the end of the WD CS of NGC 6397
in this subset of 15 orbits of GO-10424;
(2) the peak of the WD LF---what we call the end of the WD CS---of NGC
6397 is at $m_{\rm F606W}  \sim28.7$, i.e., it is $\sim 0.2$ magnitude
fainter than the peak of the M4 WD LF;
(3) we know that this is indeed the bottom of the WD CS, thanks to the
much deeper CMD obtained using the entire data base;
(4) the saturation  level of deep exposures is  at brighter magnitudes
in  NGC 6397  than  in M4,  because  of the  different exposure  times
($\sim710$ s vs. $\sim1200$ s).

The M4 field is in the core  of that cluster, while the NGC 6397 field
from GO-10424 is in the outskirts, where the bright part of the CMD is
poorly  populated.  In  order to  show the  location of  the  NGC 6397
horizontal branch on  the same CMD as the WDs, in  the middle panel of
Fig.~\ref{compa} we plot the photometry from Anderson et al.\ (2008b).
This was derived from the  same combination of camera and filters used
in GO-10424,  but with  shorter exposure times,  and with  the cluster
center in the middle of the field of view (GO-10775, PI:\ Sarajedini).

%%%%%%%%%%%%%%%%%%%%%%%%%%%%%%%%%%%%%%%%%%%%%%
%
\section{Comparison with Theory}
\label{the}
%
%%%%%%%%%%%%%%%%%%%%%%%%%%%%%%%%%%%%%%%%%%%%%%

In  this  section  we  compare  our  observed  CMD  and  LF  with  the
predictions of stellar evolution and WD cooling models.

We  have  first  fit   the  $\alpha$-enhanced  ([$\alpha$/Fe]  =  0.4)
isochrones  and  zero  age   Horizontal  Branch  (ZAHB)  sequences  by
Pietrinferni  et  al.\  (2006---hereafter  BaSTI  isochrones)  to  the
observed CMD,  from the MS to the  HB. This allows us  to estimate the
distance modulus and reddening that  will be employed for the analysis
of  the WD  cooling  sequence.   In addition,  this  fit provides  the
reference cluster  age from  the turn-off (TO)  point, to  be compared
with  the age  derived from  the WD  LF.  We  have selected  the BaSTI
$\alpha$-enhanced isochrones and ZAHB with $Z=0.004$, corresponding to
[Fe/H] = $-$1.01 (for [$\alpha$/Fe] =  0.4), a value very close to the
most   recent   spectroscopic   estimate   [Fe/H]   =   $-1.07\pm0.01$
([$\alpha$/Fe] = +0.39$\pm0.05$) by Marino et al.\ (2008).

Before using  these models to fit  the cluster CMD we  had to consider
carefully  two issues,  related to  the extinction  along the  line of
sight to this cluster.  The first point is that the extinction is high
for M4 ($A_V >$ 1), and it  also varies across the face of the cluster
(see,  e.g., Cudworth \&  Rees~1990, Liu  \& Janes~1990,  Mochejska et
al.\ 2002).   This is due  to the  fact that the  line of sight  to M4
passes through  the outer regions of the  Scorpius-Ophiucus dark cloud
complex.

It is  well known  that the  ratio between the  extinction in  a given
photometric  band  ($A_{\lambda}$)  and  $A_V$  depends  on  the  flux
distribution of the  stellar source, and is in  principle dependent on
parameters  like $T_{\rm  eff}$,  $\log g$,  and chemical  composition
(see,  e.g.,  Bedin  et  al.\  2005b  and Girardi  et  al.\  2008  for
discussions and data specific for the ACS photometric filters).
So long  as $A_V$  is small, this  effect is  also small and  a single
value  of $A_{\lambda}/A_V$  can  be safely  applied  along the  whole
isochrone.  But  the high value  of $A_V$ in  the case of M4  makes it
necessary to consider, in the  fits to the observed CMD, the variation
of $A_{\lambda}/A_V$  along the isochrones. This  effect is especially
relevant in  the F606W  band, as  shown by Fig.~2  of Girardi  et al.\
(2008) and Tables 3 and 4 of  Bedin et al.\ (2005b).  With high values
of $A_V$  one also  has to consider  the so-called  ``Forbes'' effect,
discussed in detail by Girardi  et al.\ (2008), i.e., the variation of
the  ratio $A_{\lambda}/A_V$  for a  given flux  distribution,  as the
total extinction increases.

The  second  major  issue  is  related  to  the  value  of  the  ratio
$R_V=A_V/E(B-V)$, which affects the  shape of the extinction curve.  A
number of investigations (see,  e.g., Dixon \& Longmore~1993, Ivans et
al.\ 2001) have given indications that in the direction of M4 $R_V$ is
higher than  the standard $R\sim3.1$--3.2, with values  up to $R_V\sim
4.2$  (Chini~1981). Here  we assume  $R_V=3.8$, as  in Hansen  et al.\
(2004),   consistent   with  the   value   obtained   by  Clayton   \&
Cardelli~(1998) for  $\sigma$ Sco, a star  one degree away  from M4 in
the plane of the sky.

To  include all  these  effects  in our  fits  to the  CMD  of M4,  we
proceeded as  follows.  We first  used the web interface  at {\url{\sf
http://stev.oapd.inaf.it/cgi-bin/cmd}},  which implements  the results
by Girardi et  al.\ (2008), to determine the  extinctions in the F606W
and F775W  filters, covering  the full range  of $T_{\rm eff}$  of our
isochrones  and  ZAHB,  for  $Z=0.004$,  and  for  varying  values  of
$A_V$.  These extinctions  are calculated  assuming $R_V=3.1$  and the
extinction law by Cardelli et al.\ (1989). We then included the effect
of $R_V=3.8$  by multiplying these  derived extinctions by  a constant
factor  (very close  to unity,  however) that  takes into  account the
variation   of   the   ratios  $A_{F606W}/A_V$   and   $A_{F775W}/A_V$
(calculated  at  the  central  wavelength  of the  ACS  filters)  when
changing  $R_V$ from 3.1  to 3.8,  according to  the Cardelli  et al.\
(1989) extinction law.  These final extinctions in the F606W and F775W
bands were then applied to the BaSTI isochrones transformed to the ACS
system using the bolometric corrections by Bedin et al.\ (2005b).

Figure~\ref{iso} (top-right panel) displays the fits of the MS and the
HB.
The simultaneous  match of the  ZAHB to the  lower envelope of  the HB
star distribution, and of the isochrones to the color of the unevolved
MS and red  giant branch (RGB), provide the  apparent distance modulus
$(m-M)_{F606W}=$12.68  and   the  extinction  $A_V$=1.2,   which,  for
$R_V=3.8$, corresponds to $E(B-V)=0.32$. This value is consistent with
the  mean $E(B-V)=0.33  \pm0.01$ determined  for M4  by Ivans  et al.\
(2001). Ages of  11, 12, and 13 Gyr are  displayed. They roughly cover
the full vertical  extension of the observed SGB  and match the bluest
point  along the observed  MS, hence  provide an  estimate of  the age
range  allowed by  the  isochrone-fitting procedure.   The paucity  of
stars observed in  our field along the HB phase  makes it difficult to
define unambiguously  the observed ZAHB level; we  estimated that this
adds an uncertainty of  $\pm$0.10~mag to the distance modulus obtained
from ZAHB  fitting. A $\pm$0.10~mag variation of  the distance modulus
implies an  age uncertainty  of $\sim$1~Gyr at  these old  ages, which
added in quadrature to the uncertainty at fixed distance gives a final
age estimate $t=12.0 \pm1.4$ Gyr from the isochrone fitting.

The top-left panel  of Fig.~\ref{iso} shows the fit  to the WD cooling
sequence.
We  have  employed the  same  distance,  extinction,  and ages  listed
above.  WD isochrones  for DA  objects  have been  computed using  the
Salaris et al.\  (2000) WD tracks, the same  bolometric corrections to
the ACS system as in  Bedin et al.\ (2005a), progenitor lifetimes from
BaSTI, and a default initial-final mass relationship (IFMR):
$$M_{\rm WD}=0.54 M_{\odot} \exp{[0.095(M_{\rm MS}-M_{\rm TO})]}$$
where $M_{\rm  WD}$ is the WD  mass, $M_{\rm MS}$  its progenitor mass
along  the MS,  and $M_{\rm  TO}$  is the  value of  the stellar  mass
evolving at the TO point for the given cluster age.  This relationship
has  the mathematical  form of  the IFMR  used by  Wood~(1992)  and is
{essentially the same as the one  adopted in the Hansen et al.\ (2004)
analysis, although  our constant 0.54  $M_{\odot}$ --fixed considering
the $M_{\rm WD}$ obtained from  the synthetic AGB treatment by Cordier
et al.~(2007), for $M_{\rm TO}$  values in the relevant age range-- is
slightly  different from  the  value of  0.55  $M_{\odot}$ adopted  by
Hansen et al.\ (2004).
As for the extinction corrections,  we used the Girardi et al.\ (2008)
results as described above, but we had to employ MS stars covering the
same $T_{\rm eff}$ values spanned  by our cooling sequences, since the
Girardi et al.\ (2008) calculations do not include WD models.

The  bright  part  of  the   WD  cooling  sequence,  down  to  $m_{\rm
F606W}\sim27$, appears to be reproduced very well by the models.
Some systematic color shifts appear at fainter magnitudes.  
These should  not be  interpreted in terms  of a mismatch  between the
colors  of the  faint WD  models and  the observations,  but  as large
errors in filter F775W.  
%
%%%%%%%%%%%%
Our photometry is  considerably more solid in filter  F606W (for which
we  have 20  well-dithered exposures  of  1200s each)  than in  filter
F775W. This  is due to  the heterogeneous (and limited)  material used
for this red filter, which result in larger errors.  For these reasons
we will consider the luminosity functions only in filter F606W.  [Note
that this prevents us from clearly detecting color-features at the end
of the WD CS.]
%%%%%%%%%%%%
%
We  note, however, that Kowalski (2007) shows
in  his Figure  2 two  0.6$M_{\odot}$ WDs  in the  (F606W,F814W) plane
employing   model  atmospheres   calculated  with   and   without  the
Ly-$\alpha$ far  red wing opacity  contribution (see also  Kowalski \&
Saumon, 2006).  The  effect is important only for  the cooler WDs. Our
coolest WD temperatures are around 4000 K. At this $T_{\rm eff}$, from
Kowalski's figure one can see  that F606W must vary by definitely less
that 0.1 mag, an  amount smaller than the bin size of  the LF and much
smaller  than the  typical  color error  at  these luminosities.   The
effect on  the age  from the LF  is negligible.  Kowalski  estimates a
difference of 0.5 Gyr  on NGC~6397 WD ages, but he makes  use of a fit
of the  location of the  track on the  CMD, not of  the LF, as  in our
method.

The F606W  magnitude of  the termination of  the observed  WD sequence
corresponds to age between 11 and 12 Gyr.
We also display  the location of a 12  Gyr isochrone for He-atmosphere
WDs (type DB).   These DB models have been  calculated as described in
Salaris et  al.\ (2001), with  the additional input that  the boundary
conditions  for  $T_{\rm  eff}$  below  12,000~K are  from  the  model
atmospheres by  Bergeron~(1995). The mass fraction  of the He-envelope
is  $M_{He}=10^{-3.5}M_{WD}$, the  same  value assumed  in the  Hansen
(1999) models.  Bolometric corrections to  the ACS system are from the
same  source employed  in Bedin  et al.\  (2005a), but  for  a pure-He
atmosphere.   We used  the  same  IFMR employed  to  calculate the  DA
isochrones.  The  termination of the  DB isochrone is located  at much
fainter magnitudes  than the DA case,  due to faster  cooling times at
low luminosities, and cannot be reached by the present data.

The lower  panel of Fig.~\ref{iso}  displays a comparison  between the
observed  LF in  F606W  (bin  width of  0.2~mag)  and the  theoretical
counterparts. Our primary aim here is to study the consistency between
the age  estimated from  the cooling sequence  --- which  is estimated
from the magnitude level of the  peak at the bottom of the WD sequence
--- and  the TO  age  of this  cluster.   With our  reference IFMR  we
computed theoretical LFs at fixed  age, by varying the exponent of the
progenitor  mass function,  assumed to  be a  power law  $m^{-x}$, the
percentage of DB  objects, and the percentage of  unresolved WD binary
systems (see Bedin  et al.\ 2008b for a description  of how to compute
the WD LF,  allowing for unresolved binary systems).   The position of
the  peak caused  by the  termination of  the DA  cooling  sequence is
unaffected by these parameters.  What  is changed is the height of the
peak  compared to the  level of  the bright  part of  the LF,  and the
details  of the  overall  shape of  the  LF. For  example, the  binary
systems  affect the  shape  of the  LF  mainly at  the magnitude  bins
centered around  $m_{\rm F606W}=27.7$ and  27.9, whereas the  value of
the exponent  $x$ and the  DB fraction affect  the height of  the peak
compared to the level of the LF at brighter magnitudes. In fact, there
is a certain  degree of degeneracy between the  ratio of DB/DA objects
and the exponent  of the mass function. The effect  of a mild increase
of $x$ (i.e., a more top-heavy  mass function) on the overall shape of
the LF  can be counterbalanced  by an increase  of the fraction  of DB
objects.

Assuming a DB fraction of 30\%, a typical value for the DB/DA ratio in
the disk  population (Tremblay \& Bergeron 2008),  and normalizing the
theoretical  LF to  the  observed number  of  objects between  $m_{\rm
F606W}=25$ and 26.5, the observed star counts are best reproduced by a
progenitor mass function with $x=-0.95$  (a value much higher than the
standard Salpeter $x=-2.35$ and  in remarkable agreement with what was
obtained by Bedin et al.\ 2001, $x=-1$).
A 5\%  fraction of binary WD+WD  systems improves the  fit right above
(in  brightness) the  faint-end peak  of the  WD LF,  and seems  to be
supported  by  both  observational   (Sommariva  et  al.\  2008),  and
theoretical works (Heggie \& Giersz 2008).
Theoretical  LFs  for  these  choices of  parameters  and  $(m-M)_{\rm
F606W}=12.68$ are displayed in Fig.~\ref{iso}, for the same three ages
of 11,  12, and 13~Gyr as  in the TO fits,  plus a best-fit  WD age of
11.6 Gyr, which has a $\pm$  0.6 Gyr internal (only) error, due to the
fit of the magnitude of the peak of the LF, and the uncertainty in our
derived distance modulus.

\begin{figure*}[ht!]
\epsscale{1.00}
\plotone{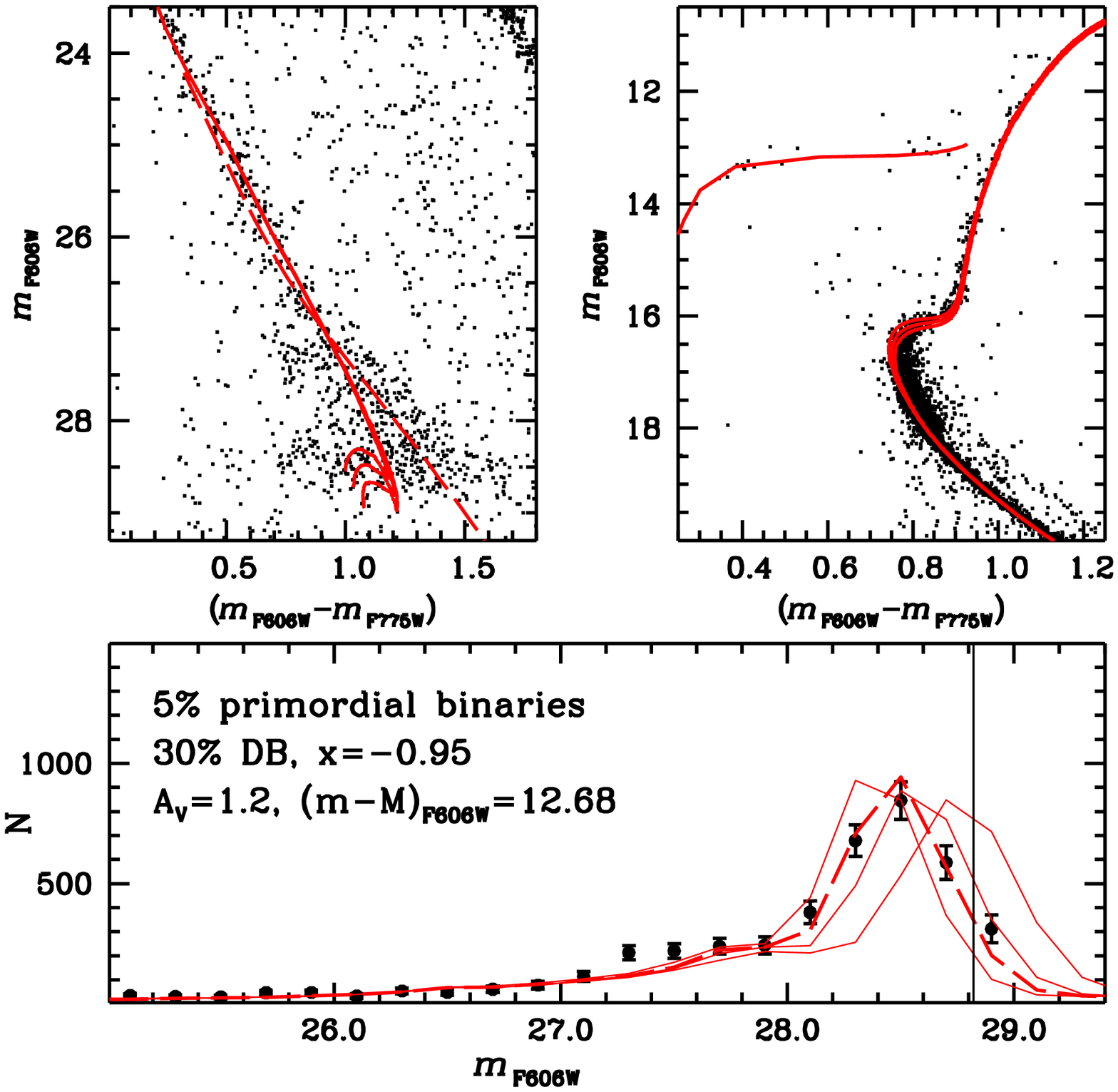}
\caption{
{\it  (upper  right panel):}  Fit  of  theoretical  isochrones to  the
cluster photometry, from the MS to the ZAHB, for ages of 11, 12 and 13
Gyr.
{\it  (upper left  panel):} Similarly  for WD  isochrones;  here solid
lines denote DA models, dashed line DB models.
The {\it lower panel} displays the fit of theoretical LFs for 11, 11.6
(dashed  line), 12,  and  13 Gyr  to  the observed  star counts.   The
parameters of  the theoretical LFs  are also displayed.  The isochrone
fits employ the  same extinction and apparent distance  modulus as the
LF fits.
}
\label{iso}
\end{figure*}

%%%%%%%%%%%%%%%%%%%%%%%%%%%
%
\section{Summary and discussion} 
%
%%%%%%%%%%%%%%%%%%%%%%%%%%%

Using only 14 $HST$ orbits, we  have been able to measure the location
of the  peak at the bottom  of the WD  CS in the globular  cluster M4.
This peak is located at magnitude $m_{\rm F606W}=28.5\pm0.1$.

We  also obtained  proper motions  for a  subsample of  our  stars and
demonstrate that most of the objects  along the observed WD CS are not
foreground or background stars in the Galactic field.
We also used a subset of the HUDF to demonstrate that our WD CS and WD
LF   are  not  appreciably   contaminated  by   unresolved  point-like
background galaxies, which our  proper motions cannot disentangle from
members, because they are too faint  to be reliably found in our first
epoch.

We have produced in a  consistent photometric system a CMD which spans
more than 18 magnitudes, going from the the bottom of the WD CS almost
to the tip  of the RGB. The  CMD from the MS to  the horizontal branch
stage has been employed to determine the distance modulus, extinction,
and  MS turn-off  age  from  isochrone fitting.  A  comparison of  the
observed  WD  LF with  results  from  theoretical  models provided  an
independent age  determination.  The  age inferred from  the WD  LF is
$11.6\pm0.6$  Gyr, in  good  (internal) agreement  with  the age  from
fitting the main sequence turn-off ($12.0\pm1.4$ Gyr).
The error bars take into  account only errors intrinsic to our methods
and to the observations,  not uncertainties in the adopted theoretical
models.   A  detailed  analysis   of  the  absolute  uncertainties  in
present-day WD models is beyond the  purpose of this work, and will be
performed in a separate paper (Salaris et al., in preparation).
Salaris~(2009) provides  a rough estimate of the  order of $\pm$2~Gyr,
that   takes   into   account   current  uncertainties   on   the   CO
stratification,  envelope composition  and thickness,  IFMR  and model
input physics.

We note  that both our  turn-off and WD  ages are consistent  with the
best-fit WD  age of 12.1 Gyr derived  for M4 by Hansen  et al.\ (2004)
employing different data, models, and methods.
%%%%%%
% 
%%%%%%
However,  Hansen et  al.\  (2004)  age determination  is  based on  an
estimate  of  distance  modulus   and  extinction  obtained  from  the
main-sequence-fitting  to a  ground-based fiducial  main  sequence, as
described in Richer  et al.\ (1997). This distance  estimate makes use
of  pre-Hipparcos  parallaxes  for  a  set of  field  subdwarfs.   The
apparent distance modulus they obtain  in $V$ is 0.17~mag shorter than
our estimate, and  their extinction $A_V$ is 0.12~mag  higher than our
result.   In addition  to  the different  pass-bands  ($V$ instead  of
$m_{\rm F606W}$) and methods, the well-documented reddening variations
across  the cluster  may  explain the  different extinction  estimates
(assuming $R_V=3.8$, a difference  in $A_V$ of 0.12~mag corresponds to
a  difference  in  reddening  of $E(B-V)\simeq$0.03  between  the  two
fields,  consistent with the  variation of  the reddening  detected in
M4).
The Hansen et  al.\ (2004) lower extinction  implies a
true distance  modulus shorter  by $\approx$0.3~mag, when  compared to
our result.  It is difficult  to fully understand the reasons for this
discrepancy.   We   can  simply  mention  some   inferences  from  the
literature.  First of all, main-sequence-fitting distances to globular
clusters that make use of Hipparcos parallaxes of field subdwarfs tend
to  be systematically  larger than  pre-Hipparcos results  (see, e.g.,
Gratton et  al.\ 1997, Carretta  et al.\ 2000).   Also, post-Hipparcos
main-sequence-fitting    distances    to    globular   clusters    are
systematically  larger  by 0.2-0.3  magnitudes  compared to  distances
obtained  from Baade-Wesslink  (BW)  methods applied  to clusters'  RR
Lyraes (Gratton et al.~1997), whereas  the result adopted by Hansen et
al.\ (2004) appears to be consistent with BW analyses of M4 RR Lyraes.
Overall, our  longer distance  to M4 seems  to be consistent  with the
globular  cluster distance scale  set by  the Hipparcos  parallaxes of
field  subdwarfs.   Our longer  distance  modulus  can  also at  least
partially  explain why  Hansen  et al.\  obtained  a WD  age of  $\sim
14-16$~Gyr with  their method and  the Salaris et al.\  (2000) cooling
models,  whereas we  obtain a  much  lower age  with the  same set  of
models.  Differences  in the method for  the WD age  estimate may also
play a  relevant role in explaining  this age discrepancy,  as well as
the fact that in our analysis  we have accounted for the effect of the
change  of the  $A_{\lambda}/A_{V}$ ratio  as a  function of  both the
stellar  flux  distribution  and  of  $A_{V}$ itself,  an  effect  not
included in Hansen et al.\ study.
%%%

To demonstrate  the robustness of  our reduction strategy,  we applied
our  techniques to  a subset  of the  ACS/WFC GO-10424  images  of the
Galactic globular cluster  NGC 6397, which we constructed  to have the
same total amount of exposure time for each filter as our M4 program.
Although conditions might  not be identical to those  of M4 (different
crowding, different red filter:\ F814W instead of F775W), we were able
to locate the peak of the WD LF  in NGC 6397 also, and found it at the
same magnitude  obtained from the entire 126-orbit  data set (Anderson
et al.\ 2008a), at $m_{\rm F606W}=28.7\pm0.1$.

It remains to be understood why the observed WD LF peak in NGC 6397 is
$\sim$0.2 magnitude fainter than in M4.
This,  of course,  can  be due  to  the effect  of different  apparent
distance  moduli  and/or  cluster  ages.   De~Angeli  et  al.\  (2005)
determined  the  relative MS  turn-off  age  for  these two  clusters,
finding that M4 is $\sim 10$\%  younger than NGC 6397.  For the M4 age
derived  in this  paper, this  implies  that NGC  6397 is  $\sim$1~Gyr
older. Such an  age difference corresponds to a WD  LF peak fainter by
$\sim$0.2~mag,  as observed.  Therefore, if  turn-off and  WD relative
ages between these two  clusters are consistent, the apparent distance
moduli have  to be approximately  the same.  According to  the results
presented in Table~3  of Richer et al.~(2008), NGC  6397 would have an
apparent distance modulus in $V$$\sim$0.1~mag larger than the estimate
by Hansen  et al.\ (2004, see  also Richer et al.\  2004, Table~1) for
M4.   Allowing for  this different  distance modulus,  the WD  LF peak
implies  a  NGC  6397  age  about  0.5~Gyr older  than  M4,  still  in
reasonable agreement with the turn-off relative ages.
We note, however, that the  difference in the apparent distance moduli
for the two clusters is of about the same size as its uncertainties.
A detailed  comparative analysis between M4 and  NGC 6397, determining
distances, reddening, TO and WD ages in a homogeneous way will be also
presented in the follow-up paper (Salaris et al., in preparation).

%%%%%%%%%%%%%%%%%%%%%%%%%%%%%%%%%%%%%%%%%%%%%%%%%%%%%%%%%%%%%%%%%%%%

\acknowledgements
We  thank  the  anonymous  referee  for the  careful  reading  of  the
manuscript, and for the many useful comments.
J.A.\ and I.R.K.\ acknowledge support from STScI grant GO-10146. 
G.P.\ acknowledges financial contribution from contract ASI-INAF I/016/07/0.

% R9: This is to have figures where I want them to be.

\newpage
~\\~ 
\newpage

\end{document}